\pdfoutput=1
\documentclass[12pt]{article}
\usepackage{amsmath}
\usepackage{amsthm}

\usepackage{graphicx,psfrag,epsf}
\usepackage{enumerate}
\usepackage{natbib}
\usepackage{url} 

\usepackage{bbm}
\usepackage{amssymb}
\usepackage{float}

\usepackage{comment}
\usepackage{soul}

\usepackage[]{algorithm2e}
 
\usepackage[T1]{fontenc}
\usepackage[latin9]{inputenc}
\usepackage{array}
\usepackage{subfig}
\usepackage{color} 	

\SetKwInput{KwInput}{Input}
\SetKwInput{KwReturn}{Return}

\newtheorem*{prop*}{Proposition}

\newcommand{\blind}{0}

\def\simiid{\,{\buildrel {\rm iid} \over \sim}\,}

\begin{document}

\def\spacingset#1{\renewcommand{\baselinestretch}%
{#1}\small\normalsize} \spacingset{2}


\if0\blind
{
  \title{\bf Regularized Estimation of Piecewise 
Constant Gaussian Graphical Models: The Group-Fused Graphical Lasso}
  \author{Alexander J. Gibberd\\
    Department of Statistical Science, University College London\\
    and \\
    James D. B. Nelson\thanks{
    In the bright memory of Dr James D. B. Nelson who sadly passed away Sep. 2016. The authors gratefully acknowledge support from the UK \textit{Defence Science \& 
Technology Laboratory (DSTL)}.}\hspace{.2cm} \\
    Department of Statistical Science, University College London}
  \maketitle
} \fi

\if1\blind
{
  \bigskip
  \bigskip
  \bigskip
  \begin{center}
    {\LARGE\bf Regularized Estimation of Piecewise 
Constant Gaussian Graphical Models: The Group-Fused Graphical Lasso}
\end{center}
  \medskip
} \fi

\bigskip
\begin{abstract}
The time-evolving precision matrix of a piecewise-constant Gaussian
graphical model encodes the dynamic conditional dependency structure of
a multivariate time-series. 
Traditionally, graphical models are estimated under the assumption that
data is drawn identically from a generating distribution. 
Introducing sparsity and sparse-difference inducing priors we relax
these assumptions and propose a novel regularized M-estimator to jointly
estimate both the graph and changepoint structure. The resulting
estimator possesses the ability to therefore favor sparse dependency
structures and/or smoothly evolving graph structures, as required. 
Moreover, our approach extends current methods to allow estimation of
changepoints that are grouped across multiple dependencies in a
system.  An efficient algorithm for estimating structure is proposed.
We study the empirical recovery properties in a synthetic setting.  The
qualitative effect of grouped changepoint estimation is then
demonstrated by applying the method on a genetic time-course data-set.

\end{abstract}

\noindent%
\vfill

\newpage
\spacingset{1.45} 

\section{Introduction}

High-dimensional correlated time series are found in many modern
socio-scientific domains such as neurology, cyber-security, genetics and
economics. A multivariate approach, where the system is modeled jointly, can
potentially reveal important inter-dependencies between variables.
However, naive approaches which permit arbitrary (graphical) dependency
structures and dynamics are infeasible because the number of possible
graphs becomes exponentially large as the number of variables increases.

For data streams with continuous-valued variables, a penalized maximum
likelihood approach offers a flexible means to estimate the underlying
dependency structure and continues to attract much attention.  In this
setting a common assumption is that the data is drawn from a
multivariate distribution where the conditional dependency structure is
in some sense sparse--- the dependency graph is expected to constitute a
small proportion of the total number of possible edges.  Typically, a
Gaussian likelihood is accompanied by a sparsity inducing prior. For
example, \cite{Ghaoui} and \cite{Friedman2008} penalize the likelihood
with an $\ell_1$ norm applied to the precision matrix (non-convex penalties have also been investigated, see \cite{Chun2014}). Further
extensions have been considered, for example \cite{Lafferty2012} who graphical model estimation in the non-parametric case, and more
recently \cite{Lee2015} study models with mixed types of
variable. 

It is of particular interest to understand how dependency graphs evolve over
time, and how prior knowledge relating to such dynamics can be exploited to
constrain the graph estimation.  Specifically, we consider how a
piecewise constant graphical model can be estimated such that the
dependency graphs are constant in locally stationary regions segmented
by a set of changepoints.  This is a challenging problem. If the
changepoints were known in advance then local graph estimation could be
performed.  However, the changepoints cannot be found without first
estimating the graphs. Previous approaches \citep{Angelosante2011} have
resorted to using dynamic programming alongside the $\ell_1$ graph
learning approaches. Unfortunately, these are restricted to quadratic
computational complexity as a function of the time-series length.  An
alternative approach as followed by \cite{Ahmed2009,Kolar2012} and others is to
formulate a convex optimization problem with suitable constraints that
encourage desired dynamical properties. We refer to such approaches as
fused graphical models.

Our contribution investigates and extends a class of models to enable the
estimation of changepoints that are grouped over multiple edges in a
graphical model. Such grouped changepoints indicate changes in
dependency across a system that influence many variables at once. In
many situations there may be a priori knowledge of potential groups (for
example, grouping over different gene function in genetic data or asset
classes in finance).  Grouped changes often indicate some sort of regime
or phase change in the system dynamics which may be of interest to the
practitioner.  To this end we propose the \emph{group-fused graphical
lasso (GFGL)} method for joint changepoint and graph estimation. We
contrast the proposed grouped estimation of changepoints in graphical
models with previous approaches which enforce changepoints at the level
of individual edges only and which therefore fail to capture such
grouping behavior.

In Section \ref{sec:DGGM}, we describe current dynamical graphical model
estimation; we introduce our main contribution in the form of the GFGL
estimator; and contrast this with previously proposed fused graphical
model approaches. Following this, in Section \ref{sec:ADMM} we present
an efficient alternating-directed method of moments (ADMM) algorithm for
estimation with GFGL. The proposed methodology is demonstrated on
simulated examples in Section \ref{sec:Synthetic}, before we consider an application looking at temporal-evolution of gene dependencies in
Section \ref{sec:Applications}. We conclude with a discussion of the
work.  Some technical details are summarized in the Appendix which can be found in the on-line supplementary material.

\section{Dynamic Gaussian Graphical Models}
\label{sec:DGGM}

Given a $P$-variate time-series
$\boldsymbol{y}^t \sim \mathcal{N} (\boldsymbol{\mu}^t,
\boldsymbol{\Sigma}^t)$ for $t=1,\ldots,T$, if the precision matrices $\boldsymbol{\Theta}^t :=
(\boldsymbol{\Sigma}^{t})^{-1}$
are well defined then the dependency structure of
the series can be captured by a dynamic Gaussian Graphical Model (GGM). This
comprises a collection of graphs $G^t = (V^t,E^t)$, where
the vertices $V^t = \{1,\ldots,P\}$
represent each component of
$\boldsymbol{y}^t$ and the edges $E^t$ represent conditional dependency
relations between variables over time. More precisely, the edge $(i,j) \in {E}^t$ is present in the graph if the $i$ and
$j$th variables are conditionally dependent given all other variables.
It follows in the Gaussian case that conditional independence $y_{i}^t \perp y_{j}^t |\:
y_{-\{i,j\}}^t$ is satisfied if and only if the corresponding entries of
the precision matrix are zero, i.e. $\Theta_{ij}^t = \Theta_{ji}^t = 0$
\citep{Lauritzen1996}. Therefore, since it encodes the conditional dependency graph, estimation
of the precision matrix is of great interest. 

Traditionally, estimation of GGM's is performed under the assumption of
stationarity, i.e. we have \emph{identically} distributed draws from a
Gaussian model.
Letting 
$
  \boldsymbol{Y}=
  (\boldsymbol{y}^1,\ldots,\boldsymbol{y}^T)
  $
be a set of observations and assuming
  $
  \boldsymbol{y}^t 
  \simiid
  \mathcal{N}(\boldsymbol{0}, \boldsymbol{\Sigma})
$, 
one can construct an estimator for $\boldsymbol{\Sigma}^{-1}$ by
maximizing the log-likelihood:
$\hat{\boldsymbol{\Theta}}:={\arg\max}_{\boldsymbol{X}}\big[\mathrm{log}\big(\mathrm{det}(\boldsymbol{X})\big)-\mathrm{tr}(\hat{ 
\boldsymbol{S}}\boldsymbol{X})\big]$, where 
$\hat{\boldsymbol{S}}=\boldsymbol{Y}\boldsymbol{Y}^{\top}/2T$. In 
the case where the number of observations is greater than the number of 
variables ($T > P$) we can test for edge significance to find a GGM \citep{Drton2004}. 
However, in the non-identical case, because only one data-point may be observed at each node per time-step 
the traditional empirical covariance estimator $\hat{\boldsymbol{S}}^t =
\boldsymbol{y}^t (\boldsymbol{y}^t)^{\top}/2$ is rank deficient for 
$P>1$.
To this end, estimation of the precision matrix requires additional
modeling assumptions. A strategy explored in several recent works
\citep{Ahmed2009, Danaher2013,Conference2014, Monti2014} is to introduce priors in the form 
of regularized M-estimators, viz. 
\begin{equation}
\hat{\boldsymbol{\Theta}}^{t} 
: = 
\underset{\boldsymbol{X}^{t} \in \{\boldsymbol{X}_{+}^{t}\}_{t = 1}^{T}}
         {\arg \min}
\big[ 
  \mathcal{L}( \{\boldsymbol{X}^{t}\} )
\big],
\label{eq:GFGL_problem}
\end{equation}
with a convex cost function: 
\begin{equation}
\mathcal{L}(\{\boldsymbol{X}^{t}\})=\sum_{t=1}^{T}-L(\boldsymbol{X}^{t},
\boldsymbol{y}^{t})+R_{\mathrm{Shrink}}(\{\boldsymbol{X}^{t}\})+R_{\mathrm{
Smooth}}(\{\boldsymbol{X}^{t}\})\;,\end{equation}
where $L(\boldsymbol{X}^{t},\boldsymbol{y}^{t})$ is proportional to
the log-likelihood and follows from the normal distribution.
The penalty terms $R_{\mathrm{Shrink}}$, $R_{\mathrm{Smooth}}$
correspond to prior shrinkage/smoothness assumptions. Typically, the
smoothness term will be a function of the difference between estimates
$\boldsymbol{X}^t-\boldsymbol{X}^{t-1}$, whereas the shrinkage term will
act at specific time points, i.e. directly on $\boldsymbol{X}^t$.

One popular approach that is relevant to GGM is to use these regularizers to place assumptions on the number of dependencies in a graph. For example, in the i.i.d. case, we need not consider conditional dependencies between all
variables, but only a small subset of those which appear most dependent.
This assumption of sparsity can be viewed as placing a prior on the
parameters $\boldsymbol{\Theta}$ to induce zeros in the off-diagonal
entries of the precision matrix.  Akin to the Laplace prior associated
with the \emph{least absolute shrinkage and selection operator (lasso)}
for linear regression \citep{Tibshirani1996}, one can construct the
\emph{graphical lasso} estimator for the precision matrix as:
$
  \hat{\boldsymbol{\Theta}} 
  := 
  \arg \min_{\boldsymbol{X}} 
  \big[
    -L(\boldsymbol{X},\hat{\boldsymbol{S}}) + \lambda\|\boldsymbol{X}\|_{1}
  \big]
$, where $L(\boldsymbol{X},\hat{\boldsymbol{S}})= \mathrm{log}\big(\mathrm{det}(\boldsymbol{X})\big)-\mathrm{tr}(\hat{ 
\boldsymbol{S}}\boldsymbol{X})$.  
This estimator, as examined by \cite{Banerjee_2007, Friedman2008} allows
one to stabilize estimation of the precision matrix in the
high-dimensional regime (where $P>T$) and estimate a sparse graph $\hat{G}$ (sparsity is controlled by $\lambda$). A full Bayesian treatment for the graphical lasso is given by \cite{Wang2012} who
investigate how representative the mode is of the full posterior.

Several approaches which incorporate dynamics in such graphical
estimators have been suggested. \cite{Zhou2010,Kolar2011} utilize a local estimate
of the covariance in the term $L(\boldsymbol{X}^t)$ by replacing
$\hat{\boldsymbol{S}}$ in the graphical lasso with a time-sensitive
weighted estimator
\begin{equation}
\hat{\boldsymbol{S}}^{t} 
= 
\sum_{s} w_{s}^{t} \boldsymbol{y}_{s}
(\boldsymbol{y}_{s})^{\top} / 
\sum_{s} w_{s}^{t} \, , 
\label{eq:kernel_covariance}
\end{equation}
where $w_{s}^{t} = K(|s-t| / h)$ are weights derived from a
symmetric non-negative smoothing kernel function $K(\cdot)$ with width
related to $h$. The resulting graphs $\hat{G}^t$ are now
representative of some temporally localized data. By making some smoothness assumptions on the underlying covariance matrix such a kernel estimator can be shown to be risk consistent \citep{Zhou2010}. \cite{Kolar2011} go further, and demonstrate that placing assumptions on the Fisher information matrix allows one to prove consistent estimation of graph structure in such dynamic GGM. In the next section we discuss how one may adapt these assumptions to estimate piecewise constant GGM.

\subsection{The group fused graphical lasso}
\label{sub:GFGL}
We propose the \emph{Group Fused Graphical Lasso (GFGL)} estimator for
estimating piecewise constant GGM. 
The model assumes data is generated at
time-point $t=1,\ldots,T$ according to
$
  \mathbb{R}^{P}
  \ni
  \boldsymbol{y}^t 
  \sim 
  \mathcal{N} (\boldsymbol{0}, \boldsymbol{\Sigma}^t)
$,
where the distribution is strictly stationary i.e. $\{\boldsymbol{\Sigma}^l=\boldsymbol{\Sigma}^m | \tau^k<l,m\le\tau^{k+1}\}$ between $k=0,\ldots,K$ changepoints (note we set $\tau^0=0,\tau^{K+1}=T$).
We propose to estimate the covariance and precision matrix at each time by minimizing a cost, as in Eq.
(\ref{eq:GFGL_problem}). The GFGL cost is given as:
\begin{equation}
\mathcal{L}(\{\boldsymbol{X}^{t}\})={\displaystyle \underset{\propto\mathrm{
-Likelihood}}{\underbrace{\sum_{t=1}^{T}\overset{-L(\{\boldsymbol{X}^t\},\{
\boldsymbol{y}^t\} ) } { 
\overbrace{\big(-\mathrm { 
log}\mathrm{det}(\boldsymbol{X}^{t})+\mathrm{tr}(\hat{\boldsymbol{S}}^{t} 
\boldsymbol{X}^{t})\big)}
}}}}\underset{\mathrm{\ell_{1}\;\mathrm{shrinkage}}}{ 
+\underbrace{\overset{R_{\mathrm{Shrink}}}{\overbrace{\lambda_{1}\sum_{t=1}^{T}
\|\boldsymbol { X } _ { -ii } ^ { t } \|_ { 1 } }}} } 
+\underset{\mathrm{group\:\ell_{2,1}\: 
smoothing}}{\underbrace{\overset{R_{\mathrm{Smooth}}}{\overbrace{\lambda_{2}{
\displaystyle 
\sum_{t=2}^{T}}\|\boldsymbol{X}_{-ii}^{t}-\boldsymbol{X}_{-ii}^{t-1}\|_{F}}}}} 
,\label{eq:GFGL_cost}\end{equation}
where $\|\boldsymbol{X}_{-ii}^t\|=\sum_{i\neq j}|X_{i,j}^t|$ is the matrix $\ell_1$ norm with the diagonal entries removed.

In the remainder of this paper we describe how one can efficiently
solve the GFGL problem and demonstrate two key properties, namely:
\begin{enumerate}
\item Estimated precision matrices encode a sparse dependency structure 
whereby many of the off axis entries are exactly zero, i.e.
$\hat{\Theta}^t_{i,j}=0$.
\item Precision matrices maintain a piecewise constant structure where 
changepoints tend to be grouped \emph{across} the precision matrix, such that
for many edges indexed by $(i,j)$ and $(l,m)$  the estimated
changepoints for the two edges are the same, viz.
$
  \hat{\mathcal{T}}_{i,j} = \hat{\mathcal{T}}_{l,m}
$ 
where 
$
  \hat{\mathcal{T}}_{i,j} = 
  \{
    \hat{\tau}^1_{ij}, \ldots, \hat{\tau}^{\hat{K}_{ij}}_{ij}
  \} 
$ 
represents the set of $\hat{K}_{ij}$ estimated changepoints $\tau^k_{ij}$ 
on the $i,j$th edge).
\end{enumerate}

\subsection{Relationship to previous proposals}
\label{sub:Relationship-to-previous}
\begin{table}[h]
\footnotesize
\begin{tabular}{|>{\centering}p{0.2\columnwidth}|>{\centering}p{0.12\columnwidth
}|>{\centering}p{0.3\columnwidth}|>{\centering}p{0.3\columnwidth}|}
\hline 
Name & References & Likelihood $L$ & Graph Smoothing 
$R_{\mathrm{Smooth}}$\tabularnewline
\hline
\hline 
Dynamic Graphical Lasso & \cite{Zhou2010} & 
$\big\{\mathrm{log}\big(\mathrm{det}(\boldsymbol{X}^{t})\big)-\mathrm{tr}(\hat{
\boldsymbol{S}}^{t}\boldsymbol{X}^{t})\big\}_{t=1}^{T}$ & via kernel 
(see Eq. \ref{eq:kernel_covariance})\tabularnewline
\hline 
Temporally smoothed $\ell_1$ logistic regression (TESLA) & \cite{Ahmed2009} & 
$\sum_{t=1}^{T}\big[\log\big(1+\exp(\boldsymbol{y}_{-i}^{t}X_{\cdot,i}^{t}
)\big)-\boldsymbol{y}_{-i}^{t}X_{\cdot,i}^{t}y_{i}^{t}\big]$ & 
$\lambda_{2}\sum_{t=2}^{T}\|\boldsymbol{X}_{-ii}^{t}-\boldsymbol{X}_{-ii}^{t-1}
\|_{1}$\tabularnewline
\hline 
Joint Graphical Lasso (JGL)* & \cite{Danaher2013} & 
$\sum_{k=1}^{K}\big[n_{k}\big(\mathrm{log}\big(\mathrm{det}(\boldsymbol{X}^{k}
)\big)-\mathrm{tr}(\hat{\boldsymbol{S}}^{k}\boldsymbol{X}^{k})\big)\big]$ & 
$\lambda_{2}\sum_{k<k'}\|\boldsymbol{X}^{k}-\boldsymbol{X}^{k'}\|_
{1}$\tabularnewline
\hline 
Fused Multiple Graphical Lasso (FMGL)* & \cite{Yang2012} & 
$\sum_{k=1}^{K}\big[n_{k}\big(\mathrm{log}\big(\mathrm{det}(\boldsymbol{X}^{k}
)\big)-\mathrm{tr}(\hat{\boldsymbol{S}}^{k}\boldsymbol{X}^{k})\big)\big]$ & 
$\lambda_{2}\sum_{k=1}^{K}\|\boldsymbol{X}^{k}-\boldsymbol{X}^{k-1}\|_{1}
$\tabularnewline
\hline 
SINGLE & \cite{Monti2014} & 
$\sum_{t=1}^{T}\big[\mathrm{log}\big(\mathrm{det}(\boldsymbol{X}^{t}
)\big)-\mathrm{tr}(\hat{\boldsymbol{S}}^{t}\boldsymbol{X}^{t})\big]$ & 
$\lambda_{2}\sum_{t=2}^{T}\|\boldsymbol{X}_{-ii}^{t}-\boldsymbol{X}_{-ii}^{t-1}
\|_{1}$\tabularnewline
\hline 
VCVS Model & \cite{Kolar2012} & For each node $p=1,\ldots,P$
$\sum_{t=1}^T \big{(} y_{t,p}-\sum_{i\ne p} y_{t,i} \beta_{i,t}\big{)}^2$ & $ \lambda_2 \sum_{t=1}^T \| \boldsymbol{\beta}_{\cdot,t}-\boldsymbol{\beta}_{\cdot,t-1} \|_2$
\tabularnewline
\hline
GFGL & (this work) & 
$\sum_{t=1}^{T}\big[\mathrm{log}\big(\mathrm{det}(\boldsymbol{X}^{t}
)\big)-\mathrm{tr}(\hat{\boldsymbol{S}}^{t}\boldsymbol{X}^{t})\big]$ & 
$\lambda_{2}\sum_{t=2}^{T}\|\boldsymbol{X}_{-ii}^{t}-\boldsymbol{X}_{-ii}^{t-1}
\|_{F}$\tabularnewline
\hline
\end{tabular}

\caption{Overview of likelihood and smoothing approaches for dynamic graphical
modeling.  Shrinkage via an $\ell_{1}$ term is common to all methods (in VCVS this is applied at the node-wise level)
above when used for edge selection. This is usually applied to off-diagonal
entries in the graph/precision matrix such that 
$R_{\mathrm{Shrink}}=\lambda_{1}\sum_{t=1}^{T}\|\boldsymbol{X}_{-ii}^{t}\|_{1}$.
* Note: these methods are not specifically designed for time-series data but for building fused models over different $k=1,\ldots,K$ classes/experiments each with $n_k$ data-points.\label{tab:graph_comparison}}
\end{table}
Unlike most previous proposals (see Table \ref{tab:graph_comparison}) GFGL penalizes changes across groups of edges in the graph. One notable exception to this can be found in the Varying-Coefficient Varying-Structure (VCVS) model of \cite{Kolar2012} who propose to select changepoints with an $\ell_2$ type norm over the differences. The motivation in that work is similar to ours. However, the authors formulate the graph-selection problem differently, utilizing a node-wise regularized regression estimator, rather than the multivariate Gaussian likelihood we use. Whilst node-wise estimation can recover the conditional dependency graph, it does not in general result in a valid (positive definite) precision matrix. This is in contrast to our approach here, where the positive-definite precision matrices can be used to define a probabilistic model via the GGM.

In particular we consider comparison to $\ell_1$ fused methods such as FMGL \citep{Yang2012}, TESLA \citep{Ahmed2009}, SINGLE
\citep{Monti2014} and JGL \citep{Danaher2013}.  These methods are similar
to each other in that they permit finding a smoothed graphical model
through a fused $\ell_{1}$ term. Throughout the paper we will refer to
models of this type as the \emph{Independent Fused Graphical Lasso
(IFGL)} with the same cost function as GFGL (see Eq.
\ref{eq:GFGL_cost}), but with the group-smoothing term replaced with an
$\ell_1$ penalized difference, such that
$
  R_{\mathrm{smooth}} 
  = 
  \lambda_2 \sum_{t=2}^T \| 
  \boldsymbol{X}^t - \boldsymbol{X}^{t-1} \|_1
$.
Rather
than focusing on the smoothly evolving graph through the kernel covariance
estimator $\hat{\boldsymbol{S}}^{t}$, we instead study the difference 
between the smoothing regularizer for IFGL
and GFGL. Throughout the rest of this paper we adopt a purely piecewise
constant graph model,
in this setting, the empirical covariance
is simply estimated with the data at time $t$ according to; 
$\hat{\boldsymbol{S}}^{t}=\boldsymbol{y}^{t}(\boldsymbol{y}^{t})^{\top}/2$.
One can think of this as using a Dirac-delta kernel for the covariance 
estimate. For example in Eq. (\ref{eq:kernel_covariance}) we can set 
$w_{s}^{t}=\delta(s-t)$.

\section{Algorithms for the group fused graphical lasso problem}
\label{sec:ADMM}

Since the penalty function of IFGL approaches solely  
comprises $\ell_1$ terms it is linearly separable. As such this
permits block-coordinate descent approaches utilized, for example, by
\cite{Friedman2008,Yang2012} whereby the precision matrix rows and columns
are sequentially updated. Unfortunately, the GFGL objective (Eq.
\ref{eq:GFGL_cost}) does not have the same linear separability
structure. This is due to the norm
$
  \|
    \boldsymbol{X}^t - \boldsymbol{X}^{t-1}
  \|
  _F
  :=
  (
    \sum_{i,j}
    (X_{i,j}^t - X_{i,j}^ {t-1})^2
  )
  ^{1/2} 
$ 
acting across the whole (or at least multiple rows/columns) of the
precision matrix. 
This lack of linear separability across the precision matrices precludes
a block-coordinate descent strategy \citep{Tseng2009}. 
Instead, we make use of the separability of the group norm (with respect
to time) and propose an \emph{Alternating Directed Method of Moments
(ADMM)} algorithm. A key innovation of our contribution is to
incorporate an iterative proximal projection step to solve  the
\emph{Group Fused Lasso} sub-problem.  Additionally, we demonstrate how
the same framework can be utilized to solve the previously proposed IFGL
problem.

\subsection{An alternating directions method of multipliers approach}

The ADMM approach we adopt to optimize the GFGL objective 
Eq. (\ref{eq:GFGL_cost})
splits $\mathcal{L}(\{\boldsymbol{X}^{t}\})$ into two separate, but
related problems. Equivalently to solving Eq. (\ref{eq:GFGL_problem})
we can solve:
\begin{eqnarray}
\hat{\boldsymbol{\Theta}}&=&\underset{\{\boldsymbol{X}^{t},\boldsymbol{Z}^{t}\}_
{ t=1 } ^ { T } } { \arg\min }  \bigg[{\displaystyle 
\sum_{t=1}^{T}\big(-\mathrm{log}\mathrm{det}(\boldsymbol{X}^{t})+\mathrm{tr}
(\boldsymbol{S}^{t}\boldsymbol{X}^{t})\big)}+\lambda_{1}\sum_{t=1}^{T}
\|\boldsymbol{Z}_{-ii}^{t}\|_{1}+\lambda_{2}{\displaystyle 
\sum_{t=2}^{T}}\|\boldsymbol{Z}_{-ii}^{t}-\boldsymbol{Z}_{-ii}^{t-1}\|_{F}\bigg]\nonumber\\
&&\mathrm{such}\;\mathrm{that}:\;\boldsymbol{X}^{t}- 
\boldsymbol{Z}^{t}=\boldsymbol{0}\;,
\label{eq:GFGL_seperable} \end{eqnarray}
where $\{\boldsymbol{X}^{t}\}$ and the auxiliary variables
$\{\boldsymbol{Z}^{t}\}$ are also constrained to be positive-semi-definite.
The augmented Lagrangian for GFGL is given
as:
\begin{eqnarray*}
\mathcal{L}(\{\boldsymbol{X}^{t}\},\{\boldsymbol{Z}^{t}\},\{\boldsymbol{Y}^{t}\}
): & = & {\displaystyle
\sum_{t=1}^{T}\big(-\mathrm{log}\mathrm{det}(\boldsymbol{X}^{t})+\mathrm{tr}
(\boldsymbol{S}^{t}\boldsymbol{X}^{t})\big)}+\lambda_{1}\sum_{t=1}^{T}
\|\boldsymbol{Z}_{-ii}^{t}\|_{1}\;\ldots\\
 & + & \lambda_{2}{\displaystyle
\sum_{t=2}^{T}}\|\boldsymbol{Z}_{-ii}^{t}-\boldsymbol{Z}_{-ii}^{t-1}\|_{F}
+\sum_{t=1}^{T}\langle\boldsymbol{Y}^{t}, \boldsymbol{X}^{t}-\boldsymbol{Z}^{t}
\rangle+\frac{\gamma}{2}\sum_{t=1}^{T} 
\|\boldsymbol{X}^{t}-\boldsymbol{Z}^{t}\|_{F}^{2},
\end{eqnarray*}
where $\{\boldsymbol{Y}^{t}\}_{t=1}^{T}$ is a set of dual matrices
$\boldsymbol{Y}^{t}\in\mathbb{R}^{P\times P}$. The difference
between ADMM and the more traditional augmented Lagrangian method
(ALM) \citep{Glowinski1989} is that we do not need to solve for 
$\{\boldsymbol{X}^{t}\}$
and $\{\boldsymbol{Z}^{t}\}$ jointly. Instead, we can take advantage
of the separability structure highlighted in Eq. (\ref{eq:GFGL_seperable})
to solve $\{\boldsymbol{X}^{t}\}$,$\{\boldsymbol{Z}^{t}\}$ separately.
By combining the inner product terms and the augmentation term we
find:
\begin{eqnarray*}
\mathcal{L}(\{\boldsymbol{X}^{t}\},\{\boldsymbol{Z}^{t}\},\{\boldsymbol{U}^{t}\}
): & = & {\displaystyle
\sum_{t=1}^{T}\big(-\mathrm{log}\mathrm{det}(\boldsymbol{X}^{t})+\mathrm{tr}
(\boldsymbol{S}^{t}\boldsymbol{X}^{t})\big)}+\lambda_{1}\sum_{t=1}^{T}
\|\boldsymbol{Z}_{-ii}^{t}\|_{1}\;\ldots\\
 & + &
\lambda_{2}{\displaystyle
\sum_{t=2}^{T}}\|\boldsymbol{Z}_{-ii}^{t}-\boldsymbol{Z}_{-ii}^{t-1}\|_{F}
+\frac{\gamma}{2}\sum_{t=1}^{T}\big(\|\boldsymbol{X}^{t}-\boldsymbol{Z}^{t}
+\boldsymbol{U}^{t}\|_{F}^{2}-\|\boldsymbol{U}^{t}\|_{F}^{2}\big),
\end{eqnarray*}
where $\boldsymbol{U}^{t}=(1/\gamma)\boldsymbol{Y}^{t}$ is a rescaled
dual variable. We write the solution at the $n$th iteration as 
$ 
  \{\boldsymbol{X}_{(n)}^{t}\} 
  = 
  \{
    \boldsymbol{X}_{(n)}^{1}, \ldots \boldsymbol{X}_{(n)}^{T}
  \}
$
and proceed by updating our estimates according to the three steps
below;
\begin{enumerate}
 \item \emph{Likelihood Update ($\mathrm{for\;}t=1,\ldots,T$):}
 \vspace{-0.1cm}
 \begin{eqnarray}
\boldsymbol{X}_{(n)}^{t} & = & 
\arg\min_{\boldsymbol{X}^{t}}\bigg[-\mathrm{log}\mathrm{det}(\boldsymbol{X}^{t}
  )+\mathrm{tr}(\hat{\boldsymbol{S}} {}^{t}\boldsymbol{X}^{t})+\frac{\gamma}{2}
\|\boldsymbol{X}^{t}-\boldsymbol{Z}_{(n-1)}^{t}+\boldsymbol{U}_{(n-1)}^{t}\|_{F}
^{2}\bigg]\;,\;
\label{eq:Likelihood_ADMM}\end{eqnarray}
\item \emph{Constraint Update:}
 \vspace{-0.1cm}
\begin{eqnarray}
\{\boldsymbol{Z}_{(n)}^{t}\} & = & 
\arg\min_{\{\boldsymbol{Z}^{t}\}}\bigg[\frac{\gamma}{2}\sum_{t=1}^{T}
\|\boldsymbol{X}_{(n)}^{t}-\boldsymbol{Z}^{t}+\boldsymbol{U}_{(n-1)}^{t}\|_{F}^{
2}+\lambda_{1}\sum\|\boldsymbol{Z}_{-ii}^{t}\|_{1}\;\ldots\nonumber \\
&  & \quad\quad\quad\ldots +\;\lambda_{2}\sum_{t=2}^{T}
\|\boldsymbol{Z}_{-ii}^{t}-\boldsymbol{Z}_{-ii}^{t-1}\|_{F}\bigg]\;,
\label{eq:Constraint_ADMM}\end{eqnarray}
\item \emph{Dual Update ($\mathrm{for\;}t=1,\ldots,T$):}
 \vspace{-0.1cm}
\begin{eqnarray}
\boldsymbol{U}_{(n)}^{t} & = & 
\boldsymbol{U}_{(n-1)}^{t}+\big(\boldsymbol{X}_{(n)}^{t}-\boldsymbol{Z}_{(n)}^{t
}\big)\;.\label{eq:Dual_ADMM}\end{eqnarray}
\end{enumerate}

\subsection{Likelihood update (Step 1)}
\label{sub:eigen}
We can solve the update for $\boldsymbol{\Theta}_{(n)}^{t}$
through an eigen-decomposition of terms in the covariance, auxiliary
and dual variables \citep{Yuan2011,Monti2014}. If we differentiate the 
objective in 
Eq. (\ref{eq:Likelihood_ADMM}) and set the result equal to zero we find:

\begin{equation}
(\boldsymbol{X}^{t})^{-1}-\gamma\boldsymbol{X}^{t}=\hat{\boldsymbol{S}}^{t}
-\gamma(\boldsymbol{Z}_{(n-1)}^{t}-\boldsymbol{U}_{(n-1)}^{t}).
\label{eq:Derivative_Likelihood}\end{equation}
Noting that $\boldsymbol{X}^{t}$ and
$\boldsymbol{S}^{t}-\gamma(\boldsymbol{Z}_{(n-1)}^{t}-\boldsymbol{U}_{(n-1)}^{t}
)$
share the same eigenvectors (see Appendix for details) we can now
solve for the eigenvalues of $\boldsymbol{X}^{t}$. For each eigenvalue
$\{x_{h}\}_{h=1}^{P}=\mathrm{eigval}(\boldsymbol{X}^{t})$ and
$\{s_{h}\}_{h=1}^{P}=\mathrm{eigval}\big(\hat{\boldsymbol{S}}^{t}
-\gamma(\boldsymbol{Z}_{(n-1)}^{t}-\boldsymbol{U}_{(n-1)}^{t})\big)$
we can construct the quadratic equation $x_{h}^{-1}-\gamma x_{h}=s_{h}$.
The right hand side of Eq. (\ref{eq:Derivative_Likelihood}) contains
evidence from the data-set via $\hat{\boldsymbol{S}}^{t}$, but also
takes into account the effect our priors encoded in
$\boldsymbol{Z}_{(n-1)}^{t}$,
from the non-smooth portion of Eq. (\ref{eq:GFGL_seperable}).
Upon solving for $x_{h}$ given $s_{h}$ we find:
\[
x_{h}=\frac{1}{2\gamma}\big(-s_{h}+\sqrt{s_{h}^{2}+4\gamma}\big)\;.\]
The full precision matrix $\boldsymbol{X}^{t}$ can now be found through
the eigen-decomposition:
\[
\boldsymbol{X}_{(n)}^{t}=\boldsymbol{V}\boldsymbol{Q}\boldsymbol{V}^{\top},\]
where $\boldsymbol{V}$ contains the eigenvectors of
$\hat{\boldsymbol{S}}^{t}-\gamma(\boldsymbol{Z}_{(n-1)}^{t}-\boldsymbol{U}_{
(n-1)}^{t})$
as columns and $\boldsymbol{Q}\in\mathbb{R}^{P\times P}$ is a diagonal
matrix populated by the eigenvalues $x_{h}$, ie $Q_{hh}=x_{h}$.  We note
that, by choosing the positive solution for the quadratic, we ensure
that $\boldsymbol{X}_{(n)}^{t}$ is positive-definite and thus produces a
valid estimator for the precision matrix. Since Eq.
(\ref{eq:Likelihood_ADMM}) refers to an estimation at each time-point
separately, we can solve for each $\boldsymbol{X}_{(n)}^{t}$
independently for $t={1,\ldots,T}$ to yield the set
$\{\boldsymbol{X}_{(n)}^{t}\}_{t=1}^{T}$. Indeed this update can be
computed in parallel, as appropriate.

\subsection{Group fused lasso signal approximator (Step 2)}
\label{sub:gflsa}
The main difference between this work and previous approaches is in the
use of a grouped constraint. This becomes a significant challenge when
updating $\{\boldsymbol{Z}^{t}\}$ in Eq. (\ref{eq:Constraint_ADMM}). 
Unlike the calculation of $\{\boldsymbol{X}_{(n)}^{t}\}$, we cannot
separate the optimization over each time-step.  Instead, we must solve
for the whole set of matrices $\{\boldsymbol{Z}^{t}\}$ jointly. In
addition, due to the grouped term in GFGL, we cannot separate the
optimization across individual edges.  
In contrast to independent penalization strategies \citep{Monti2014,
Danaher2013} it is not possible to solve GFGL for $\{ X_{ij}^{t}
\}$ independently of $\{ X_{kl}^{t} \}$, where $(i, j) \neq (k, l)$.
Such an inconvenience is to be expected as the constraints which extract
changepoints in GFGL can act across all elements in
$\boldsymbol{X}^t$.

For notational convenience we re-write step two in vector form. Since
each $\boldsymbol{Z}^{t}$ is symmetric about the diagonal we can reduce
the number of elements by simply taking the elements above the diagonal
$\boldsymbol{z}^{t}=(Z_{i,j}^{t}|\:\mathrm{for\:}j>i,\;i=1,\ldots,P)^{\top}$.
We then construct a matrix form such that
$\boldsymbol{Z}=(\boldsymbol{z}^{1},\ldots,\boldsymbol{z}^{T})^{\top}\in\mathbb{
R}^{T\times
P(P-1)/2}$, whereby row $t$ of the matrix correspond to values at
time-step $t$. We perform similar transformations for
$\boldsymbol{X}^{t}\rightarrow\boldsymbol{X}$ and
$\boldsymbol{U}^{t}\rightarrow\boldsymbol{U}$, and set $\bar{\lambda}_{1}=\lambda_{1}/\gamma$
and $\bar{\lambda}_{2}=\lambda_2/\gamma$ \footnote{Note that, since we have
essentially split the data in half (due to symmetry), we may wish to adjust the
lambdas to be consistent with the original problem specification in Eq.
(\ref{eq:GFGL_cost}).}. Re-writing the objective in Eq.
(\ref{eq:Constraint_ADMM}) with these transformations yields the cost function

\begin{equation}
G(\boldsymbol{Z};\bar{\lambda_{1}},\bar{\lambda_{2}})=\underset{L(\boldsymbol{Z})}{
\underbrace{\frac{1}{2}\|\boldsymbol{X}_{(n)}-\boldsymbol{Z}+\boldsymbol{U}_{ 
(n-1)}\|_{F}^{2}}}+\underset{R_{1}(\boldsymbol{Z})}{\underbrace{\bar{\lambda}_{
1 
}\|\boldsymbol{Z}\|_{1}}}+\underset{R_{2}(\boldsymbol{Z})}{\underbrace{\bar{ 
\lambda}_{2}\|\boldsymbol{D}\boldsymbol{Z}\|_{2,1}}},\label{eq:step2_G}
\end{equation}
where $\boldsymbol{D}\in\mathbb{R}^{(T-1)\times T}$ is a backwards
differencing matrix of the form $D_{i,i}=-1,D_{i,i+1}=-1$ for $i=1,\ldots,T-1$ and zero otherwise, the the group $\ell_{2,1}$ norm is defined as $\|\boldsymbol{X}\|_{2,1}:=\sum_{t}\|X_{t,\cdot}\|_{2}$. If one constructs a target matrix
$\boldsymbol{A}=\boldsymbol{X}_{(n)}+\boldsymbol{U}_{(n-1)}$ then

\begin{equation}
\mathcal{Z}(\boldsymbol{A};\bar{\lambda_{1}},\bar{\lambda_{2}})
=\arg\min_{\boldsymbol{Z}}
G(\boldsymbol{Z};\bar{\lambda_{1}},\bar{\lambda_{2}}),\label{eq:prox_GFLSA}
\end{equation}
looks like a signal approximation problem, we will refer to this problem
as the \emph{Group-Fused Lasso Signal Approximator (GFLSA)}. This
looks similar to the previously studied \emph{Fused Lasso Signal Approximator 
(FLSA)}
\citep{Liu2010} but crucially $R_2(\boldsymbol{Z})$ incorporates a group $\ell_{2,1}$ 
norm rather than the $\ell_{1}$ norm of FLSA. 

We note Eq. (\ref{eq:prox_GFLSA}) can also be thought of as a
proximity operator, such that 
$\mathcal{Z}(\boldsymbol{A};\lambda_{1},\lambda_{2})
\equiv\mathrm{prox}_{R_{1}+R_{2}}(\boldsymbol{A})$.
If $R_{1}$ and $R_{2}$ were indicator functions of two closed convex
sets $C$ and $D$ respectively, then
$\mathcal{Z}(\boldsymbol{A};\lambda_{1},\lambda_{2})$ would find the
best approximation to $\boldsymbol{A}$ restricted to the set $C\cap D$.
Unlike FLSA which penalizes the columns of $\boldsymbol{Z}$
independently, we find
$\mathrm{prox}_{R1+R2}(\boldsymbol{A})\neq\mathrm{prox}_{R2}\big(\mathrm{
prox}_{ R1}(\boldsymbol{A})\big)$ and cannot apply the two-stage
smooth-then-sparsify theorem of \cite{Friedman_2007b, Liu2010}.
Instead, we follow the work of \cite{Ala2013} and adopt an iterative
projection approach which utilizes Dykstra's method \citep{Combettesa}
to find a feasible solution for both the group fused $\ell_{2,1}$ and
lasso $\ell_{1}$ constraints. 

For any unconstrained optimal point
$\boldsymbol{Z}^{*}=\arg\min_{\boldsymbol{Z}}L(\boldsymbol{Z})$ there
exists a set of parameters $(\lambda_{1},\lambda_{2})\in[0,\infty)$ which
  will act to move the optimal point of the regularized case
  $\boldsymbol{Z}_{r}^{*}=\arg\min_{\boldsymbol{Z}}G(\boldsymbol{Z}
  ;\lambda_{1},\lambda_{2})$ such that
  $\boldsymbol{Z}^{*}\ne\boldsymbol{Z}_{r}^{*}$ where:
\[ \boldsymbol{Z}_{r}^{*} =  \underset{\boldsymbol{Z}}{\arg\min}
  L(\boldsymbol{Z}), \quad \text{ subject to } \quad
  \|\boldsymbol{Z}\|_{1} \le l_{1} \quad
  \mathrm{and}\quad\sum_{t=2}^{T}\|(\boldsymbol{DZ})_{t,\cdot}\|_{2}\le
l_{2}.  \]
For a given likelihood term, we can obtain an $l_{1}$ sparse and $l_{2}$
smooth solution by solving a penalized problem instead of the explicitly
constrained version above. Such a penalized form is found in Eq.
(\ref{eq:step2_G}) and, while $R_{1}(\lambda_{1},\boldsymbol{Z})$ and
$R_{2}(\lambda_{2},\boldsymbol{Z})$ are not explicitly indicator
functions (i.e. they do not take values $\infty$ outside some feasible
region), there does exist a mapping
between the values of the parameters
$\lambda_{1}\geq0\:,\lambda_{2}\ge0$ and the corresponding $l_{1},l_{2}$
sparsity and smoothness constraints. To give some intuition, for a given
constraint level $l_1$ and function $L(\boldsymbol{Z})$, the size of the
feasible set given by
$C_{\lambda_{1}}=\{\boldsymbol{Z}\:|\:\lambda_{1}\|\boldsymbol{Z}\|_{1}\le
l_{1}\}$, reduces as $\lambda_1$ increases. 
Thus sparsity is a monotonically non-decreasing function of
$\lambda_{1}$
The same argument can be constructed for smoothing and the constraint
set
$
  D_{\lambda_{2}} = 
  \{ 
    \boldsymbol{Z}\:|\: \lambda_{2} \sum_{t} 
    \| (\boldsymbol{DZ})_{t, \cdot} \|_{2} \le l_{2}
  \}
$. 
The proximal Dykstra method provides a way to calculate a
point $\boldsymbol{Z}_{r}^{*}\in C_{\lambda_{1}}\cap D_{\lambda_{2}}$
that is, in the sense of the $\ell_{2}$ distance, \emph{close} or
\emph{proximal} to the unconstrained solution for
$\arg\min_{\boldsymbol{Z}}L(\boldsymbol{Z})=\boldsymbol{A}$.  By
iterating between the feasibility of a solution in $C_{\lambda_{1}}$ and
$D_{\lambda_{2}}$ (see Algorithm
\ref{alg:Dykstras-iterative-projection}), a solution can be found which
is both suitably smooth and sparse.

\begin{algorithm}[H]  
\KwResult{$\mathrm{prox}_{R_1+R_2}(\boldsymbol{A})$}
$\boldsymbol{Z}_{(0)}=\boldsymbol{A}\; ,
\boldsymbol{U}_{(n)}=\boldsymbol{0}\;,\;\boldsymbol{Q}_{(n)}=\boldsymbol{0}$\\ 
 
\While{not converged,$\;n=0,1,\ldots$}{     
$\boldsymbol{V}_{(n)}=\mathrm{prox}_{R2}(\boldsymbol{Z}_{(n)}+\boldsymbol{U}_{
(n)})$\\    
$\boldsymbol{U}_{(n+1)}=\boldsymbol{Z}_{(n)}+\boldsymbol{U}_{(n)}-\boldsymbol{V}
_{(n)}$\\    
$\boldsymbol{Z}_{(n+1)}=\mathrm{prox}_{R1}(\boldsymbol{V}_{(n)}+\boldsymbol{Q}_{
(n)})$\\    
$\boldsymbol{Q}_{(n+1)}=\boldsymbol{V}_{(n)}+\boldsymbol{Q}_{(n)}-\boldsymbol{Z}
_{(n+1)}$\\   
}
\caption{Dykstras iterative projection algorithm 
\label{alg:Dykstras-iterative-projection}}
\end{algorithm}

Given that iterative projection can be used to find a feasible point,
the challenge is now to compute the separate proximity operators for
$R_1$ and $R_2$. The proximal operator for the $\ell_{1}$ term
$\mathrm{prox}_{R1}(\boldsymbol{A})$ is given by the
\emph{soft-thresholding} operator \citep{Tibshirani1996}:
\begin{eqnarray}
\mathrm{prox}_{R1}(\boldsymbol{A};\lambda_{1}) & = & 
\underset{\boldsymbol{Z}}{\arg\min}\frac{1}{2}\|\boldsymbol{Z}-\boldsymbol{A}\|_
{F}^{2}+\lambda_{1}\|\boldsymbol{Z}\|_{1}\nonumber\\
 & = & \mathrm{sign}(\boldsymbol{A})\odot\max(|\boldsymbol{A}|-\lambda_{1}, 
\boldsymbol{0})\;,\label{eq:soft-thresh}\end{eqnarray}
where the $\max$ and $\mathrm{sign}$ functions act in an element-wise
manner and $\odot$ denotes element-wise multiplication. 

Computing the group-fused term $\mathrm{prox}_{R2}(\boldsymbol{A})$
is more involved and there is no obvious closed-form solution, instead
we tackle this through a block-coordinate descent approach similar
to that considered by \cite{Bleakley} and \cite{Yuan_2005}. Our
target here is to find the proximal operator for the group smoothing
aspect of the regularizer, which we write as:
\begin{equation}
\mathrm{prox}_{R2} (\boldsymbol{A};\lambda_{2}) 
=
\underset{\boldsymbol{Z}}{\arg\min} 
\frac{1}{2}
\| \boldsymbol{Z} - \boldsymbol{A} \|_{F}^{2} 
+
\lambda_{2}
\|\boldsymbol{D} \boldsymbol{Z}\|_{2,1}.
\end{equation}
Re-writing the above 
with $\boldsymbol{\Omega} = \boldsymbol{D} \boldsymbol{Z}$ and 
constructing $\boldsymbol{Z}$ as a sum of differences via
$Z_{t,\cdot} = \boldsymbol{\omega} + \sum_{i=1}^{t-1} \Omega_{i,\cdot}$,
(where $\boldsymbol{\omega}=Z_{1,\cdot}$) then one can interpret the
proximal operator as a group lasso problem \citep{Bleakley} . Writing the 
re-parameterized
problem in matrix form one can show that solving for the jump parameters
allows us to reconstruct an estimate for $\boldsymbol{Z}$. This is formally 
equivalent to a group lasso
\citep{Yuan_2005} class of problem:

\begin{equation}
\hat{\boldsymbol{\Omega}}:=\underset{\boldsymbol{\Omega}\in\mathbb{R}^{
(T-1)\times P(P-1)/2}}{ 
\arg\min}\frac{1}{2}\|\bar{\boldsymbol{A}}-\bar{\boldsymbol{R}}\boldsymbol{
\Omega}\|_{F}^{2}+\lambda_{2}\|\boldsymbol{\Omega}\|_{2,1}\;,
\label{eq:group_lasso}
\end{equation}
where a bar $\bar{\boldsymbol{X}}$ denotes a column centered matrix and
$\boldsymbol{R}\in\mathbb{R}^{T\times(T-1)}$ is a matrix with entries
$R_{i,j}=1$ for $i > j$ and $0$ otherwise. The problem above can be
solved through a block-coordinate descent strategy, sequentially
updating the solution for each block $\Omega_{t,\cdot}$ for
$t=1,\ldots,T-1$ (see Appendix).  We can then construct a solution for
$\hat{\boldsymbol{Z}}$ by summing the differences and noting that
the optimal value for $\boldsymbol{\omega}$ is given by
$\hat{\boldsymbol{\omega}}=\boldsymbol{1}_{1,T}(\boldsymbol{A}-\boldsymbol{R}
\hat{\boldsymbol{\Omega}})$.
Correspondingly, the proximal operator for $R_2$ is found via
\begin{equation}
\mathrm{prox}_{R2}(\boldsymbol{A};\lambda_{2})=\big(\hat{\boldsymbol{\omega}}^{
\top},(\hat{\boldsymbol{\omega}}+\hat{\Omega}_{1,\cdot})^{\top},\ldots,(\hat{
\boldsymbol{\omega}}+\sum_{i=1}^{T-1}\hat{\Omega}_{i,\cdot})^{\top}\big)
{}^{\top}.\end{equation}
The overall subproblem Eq. (\ref{eq:step2_G}) can now be solved through
iteratively applying the proximity operators according to Dykstra's
algorithm (Alg. \ref{alg:Dykstras-iterative-projection}).

\subsection{Dual update and convergence (step 3)}
The final step in the ADMM-based method is to update the dual variable
via Eq. (\ref{eq:Dual_ADMM}). 
Convergence properties of general ADMM algorithms are analyzed in 
\cite{Glowinski1989}. The sequence of solutions 
$\{\boldsymbol{X}_{(n)}\}_{n\in\mathbb{N}}$
can be shown to converge \citep{Eckstein1992} to the solution of the problem:
$
\arg\min_{\boldsymbol{X}\in\mathbb{R}^{N}} 
\; 
f(\boldsymbol{X}) + g(\boldsymbol{L}\boldsymbol{X})
$,
under conditions that $\boldsymbol{L}^{\top}\boldsymbol{L}$ is invertible
and the intersection between relative interiors of domains is non-empty: 
$
  (\mathrm{ri}\:\mathrm{dom}\: g) \cap 
   \mathrm{ri}\; \boldsymbol{L} (\mathrm{dom}\: f)
   \neq \emptyset
$. 
In the GFGL and IFGL problems considered here one simply sets
$\boldsymbol{L = \boldsymbol{I}}$, in order to restrict $\boldsymbol{X}
= \boldsymbol{Z}$. Clearly in this case $\boldsymbol{I}^{\top}
\boldsymbol{I}$ is invertible and $\mathrm{dom}\:g =
\boldsymbol{I}(\mathrm{dom}\:f)$; thus the relative interiors
intersect.

Whilst ADMM is guaranteed to converge to an optimal solution, in
practice it converges relatively fast to a useful solution, but
very slowly if high accuracy is required. Following the approach of
\cite{Boyd2011} we consider tracking two convergence criteria: one
tracking \emph{primal feasibility}:
$r_{\mathrm{prime}}=\sum_{t=1}^{T}\|\boldsymbol{X}_{(n)}^{t}
-\boldsymbol{Z}_{(n) }^{t}\|_{F}^{2}\;$, relating to the optimality
requirement $\boldsymbol{X}^{*}-\boldsymbol{Z}^{*}=\boldsymbol{0}$, and
the other looking at \emph{dual
feasibility}: $r_{\mathrm{dual}}=\sum_{t=1}^{T}\|\boldsymbol{Z}_{(n)}^{t}
-\boldsymbol{Z}_{ (n-1)}^{t}\|_{F}^{2}\;$, which tracks the requirement
that $\boldsymbol{0}\in\nabla f(\boldsymbol{X}^{*})+\boldsymbol{U}^{*}$, where 
$^*$ denotes optimal value.  The rate at which the algorithm
converges is somewhat tunable through the $\gamma$ parameter. However it
is not clear how to find an optimal $\gamma$ for a given problem. In
practice we find that a value of order $\gamma=10$ provides reasonably
fast convergence which with tolerances order;
$r_{\mathrm{prime}}<\epsilon_{\mathrm{prime}}=10^{-3}$ and
$r_{\mathrm{dual}}<\epsilon_{\mathrm{dual}}=10^{-3}$.

\begin{algorithm}[!h]
\KwData{$\boldsymbol{y}^1,\ldots,\boldsymbol{y}^T$}
\KwInput{$\lambda_1,\;\lambda_2,\;\gamma,\;\epsilon_{\mathrm{dual}},\;\epsilon_{
\mathrm{prime}} $}
\KwResult{$\{\hat{\boldsymbol{\Theta}}^1,\ldots, \hat{\boldsymbol{\Theta}}^T\} 
$}
Calculate covariance matrix: 
$\hat{\boldsymbol{S}}^t=\boldsymbol{y}^t(\boldsymbol{y}^t)^{\top}/2\;\mathrm{for
}\;t=1,\ldots,T$\\
Initialize: 
$\boldsymbol{Z}^t_{(0)}=\boldsymbol{X}^t_{(0)}=\boldsymbol{U}^t_{(0)}
=\boldsymbol{0}$\\  
\While{not converged 
($r_{\mathrm{prime}}\ge\epsilon_{\mathrm{prime}},\;r_{\mathrm{dual}}\ge\epsilon_
{\mathrm{dual}}$), $\;n=0,1,\ldots$}{   
\For{t=1,\ldots,T}{
Eigen-decomposition: 
$\{s_{h},\;\boldsymbol{v}_{h}\}_{h=1}^{P}=\mathrm{eigen}\big(\hat{\boldsymbol{S}
}^{t}-\gamma(\boldsymbol{Z}_{(n-1)}^{t}-\boldsymbol{U}_{(n-1)}^{t})\big)$\\  
$x_{h}=\big(-s_{h}+\sqrt{s_{h}^{2}+4\gamma}\big)/2\gamma\;$\\
$\boldsymbol{V}=(\boldsymbol{v}_1,\ldots.\boldsymbol{v}_P),\;\boldsymbol{Q}
=\mathrm{diag}(x_1,\ldots,x_P)$\\
Apply constraints: 
$\boldsymbol{X}_{(n)}^{t}=\boldsymbol{V}\boldsymbol{Q}\boldsymbol{V}^{\top}$\\
}
$\boldsymbol{Z}_{(n)}=\mathrm{prox}_{R1+R2}(\boldsymbol{X}_{(n)}+\boldsymbol{U}_
{(n-1)}\;;\lambda_1/\gamma,\lambda_2/\gamma)\;$ // \emph{GFLSA} via Dykstras 
method* \\
$\boldsymbol{U}_{(n)}^{t}=\boldsymbol{U}_{(n-1)}^{t}+\big(\boldsymbol{X}_{(n)}^{
t}-\boldsymbol{Z}_{(n)}^{t}\big),\;\mathrm{for\;}t=1,\ldots,T$\\
$r_{\mathrm{prime}}=\sum_{t=1}^{T}\|\boldsymbol{X}_{(n)}^{t}-\boldsymbol{Z}_{(n)
}^{t}\|_{F}^{2},\;r_{\mathrm{dual}}=\sum_{t=1}^{T}\|\boldsymbol{Z}_{(n)}^{t}
-\boldsymbol{Z}_{(n-1)}^{t}\|_{F}^{2}$
}
\KwReturn{$\{\hat{\boldsymbol{\Theta}}^t=\boldsymbol{X}^t,\ldots\}$}
\vspace{0.3cm}
\caption{Outline of ADMM algorithm for GFGL. Note to solve IFGL we simply 
replace the update (*) with 
$\boldsymbol{Z}_{(n)}=\mathrm{prox}_{R1+R3}(\boldsymbol{X}_{(n)}+\boldsymbol{U}_
{(n-1)}\;;\lambda_1/\gamma,\lambda_2/\gamma)\;$ which can be computed through 
the sub-gradient finding 
algorithm as proposed in \cite{Liu2010}.}
\label{al:ADMM}
\end{algorithm}

At this point it is worth noting that there are a variety of ways one can break down Eq. (\ref{eq:GFGL_cost}) as an ADMM problem. In this paper we proceed by simply adding one set of auxiliary variables $\boldsymbol{Z}$ as in Eq. (\ref{eq:GFGL_seperable}); however, one could also adopt a \emph{linearized} ADMM scheme \citep{Parikh2013} to deal with the differencing total-variation term. A linearized scheme would result in a different set of problems for the proximity updates. The motivation for splitting the problem up as we have, constraining $\boldsymbol{X}^t=\boldsymbol{Z}^t$, is that in the IFGL case we can solve the constraint update (c.f. Eq. \ref{eq:Constraint_ADMM}) using the efficient fused lasso signal approximator algorithm of \cite{Liu2010}. Given that we are interested in how the solution of the GFGL and IFGL estimators compare it is prudent to ensure that the formulation of the algorithm is similar for both objectives. For example, given our formulation, we know at each step $\boldsymbol{Z}$ will be exactly sparse and the augmented weighting $\gamma$ is comparable for both IFGL and GFGL.

\subsection{A solver for the Independent Fused Graphical Lasso}
\label{sub:A-solver-for}

The main comparison in this paper is between the GFGL and the IFGL
classes of estimators that, respectively, fuse edges on an group and
individual level.  It is worth noting that the ADMM (Algorithm
\ref{al:ADMM}) described for GFGL can easily be adapted for such IFGL
problems by modifying the second step that corresponds to the non-smooth
constraint projection. In place of Eq. (\ref{eq:step2_G}), we construct
a fused lasso problem:
\begin{equation}
G(\boldsymbol{Z};\bar{\lambda_{1}},\bar{\lambda_{2}})=\underset{L(\boldsymbol{Z})}{
\underbrace{\frac{1}{2}\|\boldsymbol{A}-\boldsymbol{Z}\|_{F}^{2}}}+\underset{R_{
1}(\boldsymbol{Z})}{\underbrace{\bar{\lambda}_{1}\|\boldsymbol{Z}\|_{1}}}
+\underset{R_{3}(\boldsymbol{Z})}{\underbrace{\bar{\lambda}_{2}\|\boldsymbol{D}
\boldsymbol{Z}\|_{1}}\;,}
\end{equation}
where $\boldsymbol{A}=\boldsymbol{X}_{(n)}+\boldsymbol{U}_{(n-1)}$
and we replace the $\|\cdot\|_{2,1}$ norm of GFGL with a simple $\ell_{1}$
penalty of IFGL. Since the $\ell_{1}$ norm
is linearly separable, i.e. $\|\boldsymbol{X}\|_{1}=\sum_{ij}|X_{ij}|$, the
objective can now be viewed as a series of $P(P-1)/2$
separate FLSA problems. This can be solved
efficiently with gradient descent. In the IFGL case there is
no need to apply the iterative Dykstra projection as one can show the proximity
operator can be calculated as
$\mathrm{prox}_{R1+R3}(\boldsymbol{A})=\mathrm{prox}_{R3}\big(\mathrm{prox}_{R1}
(\boldsymbol{A})\big)$
\citep{Liu2010}.

\section{Synthetic Experiments}
\label{sec:Synthetic}

IFGL and GFGL are here applied to simulated, piecewise stationary, multivariate
time-series data.  This provides a numerical comparison of
their relative abilities to (i) recover the graphical structure and (ii)
detect changepoints.

\subsection{Data simulation}
\label{sub:sim}
To validate the graphical recovery performance of the estimators, data
is simulated according to a known ground truth set of precision matrices
$\{\boldsymbol{\Theta}^{t}\}_{t=1}^T$.  The simulation is carried out such that,
for a given number $K^{*}$ of ground truth changepoints $\mathcal{T}^{*}
= \{\tau_{1}, \ldots, \tau_{K^{*}}\}$, there are $K^{*}+1$ corresponding
graph structures. 
For each segment $k=1,\ldots,K^*+1$, graphical structure is
simulated uniformly at random from the set of graphs with vertex size
$|V_k|=P$ and $|E_k|=M_k$ edges, i.e. $G(V,E_{k})\sim
\mathrm{Erd\ddot{o}sR\acute{e}nyi}(P,M_k)$. A draw of $G(V,E_{k})$ can then be
used to construct a valid GGM by equating the sparsity pattern of the adjacency
matrix and precision matrix, i.e. $(i,j)\in E_k \iff \Theta^{(k)}_{i,j}\neq
0$. 

Precision matrices are formed by taking a weighted identity matrix
$ \frac{1}{2} \boldsymbol{I} \in \mathbb{R}^{P\times P} $ 
and inserting off-diagonal elements according to edges $E_k$ that are uniformly
weighted in the range $[-1,-1/2] \cup [1/2,1]$.
The absolute value of these elements is then added to the appropriate diagonal
entries to ensure positive semi-definiteness. To focus on the study of
correlation structure between variables, the variance of the distributions are
normalized such that
$(\Theta^t_{ii})^{-1} = 1$ for $i=1,\ldots,P$.

\subsection{Hyper-parameter selection}

With most statistical estimation problems there are a set of associated tuning
parameters (common examples include; kernel width/shape, window sizes, etc.) 
which must be specified. In the GFGL and IFGL model, one 
can consider the regularizer terms $R_\mathrm{Shrink}(\lambda_1)$
and $R_{\mathrm{Smooth}}(\lambda_2)$ in Eq.
(\ref{eq:GFGL_cost}) as effecting prior knowledge on the model 
parameterization.
Given this viewpoint, selection of tuning parameters $(\lambda_1, \lambda_2)$
corresponds to specification of hyper-parameters for graph sparsity and 
smoothing.

The recovery performance  will depend on the strength of priors employed.
As such $\lambda_1$ and $\lambda_2$ must be tuned, or otherwise estimated,
such that they are appropriate for a given data-set or task. In comparison to
models
which utilize only one regularizing term (for example, the graphical lasso of
\cite{Banerjee_2007}) the potential interplay between
$R_\mathrm{Shrink}(\lambda_1)$ and $R_{\mathrm{Smooth}}(\lambda_2)$ sometimes conflates the
interpretation of the different regularizers. For example, whilst $\lambda_1$
predominantly effects the sparsity of the extracted graphs, $\lambda_2$ can also
have an
implicit effect through smoothing (see Appendix for more details). 

In the synthetic data-setting, the availability of ground-truth or labeled data
affords the
opportunity to learn the hyper-parameters via a supervised scheme.
 In order to avoid repeated use of data, the simulations are split into
test and training groups which share the same ground-truth structure
$\{\boldsymbol{\Theta}^1,\ldots, \boldsymbol{\Theta}^T\}$, but are
independently sampled.
The IFGL and GFGL problems are then solved for each pair of parameters
$(\lambda_{1},\lambda_{2})$ over a search grid.
Optimal hyper-parameters can then be selected according to a
relevant measure of performance. Typically \citep{Zhou2010} one considers either
predictive risk (approximation of the true distribution), or model recovery
(estimation of the correct sparsity pattern).
In addition to tuning parameters via cross-validation, we also compared this to estimation via heuristics such as BIC. 
However, when applied to the IFGL/GFGL methods in this work such heuristics resulted in poor graph recovery performance (see Appendix for more details).

\subsection{Model recovery performance}

 Considering the model recovery setting, the problem of selecting edges can be 
treated as a binary classification problem. One popular measure of performance 
for such problems is the $F_{\beta}$-score
\begin{equation}
F_{\beta} = \frac{(1+\beta^{2})TP} 
                 {(1+\beta^{2})TP + 
                  \beta^{2}FN+FP} 
\;,
\end{equation}
where $TP$ considers the number of correctly classified edges, whilst
$FP$ and $FN$ relate to the number of \emph{false positives} and
\emph{false negatives} (Type 1 and Type 2 errors) respectively (a score of 
$F_{\beta}=1$ represents perfect recovery). Since
dynamic network recovery is of interest, the average
$F_1$-score
is taken over each time-series to measure the
effectiveness of edge selection.
\begin{figure}[h]
\centering{
\subfloat[]{
\includegraphics[width=0.5\columnwidth]
{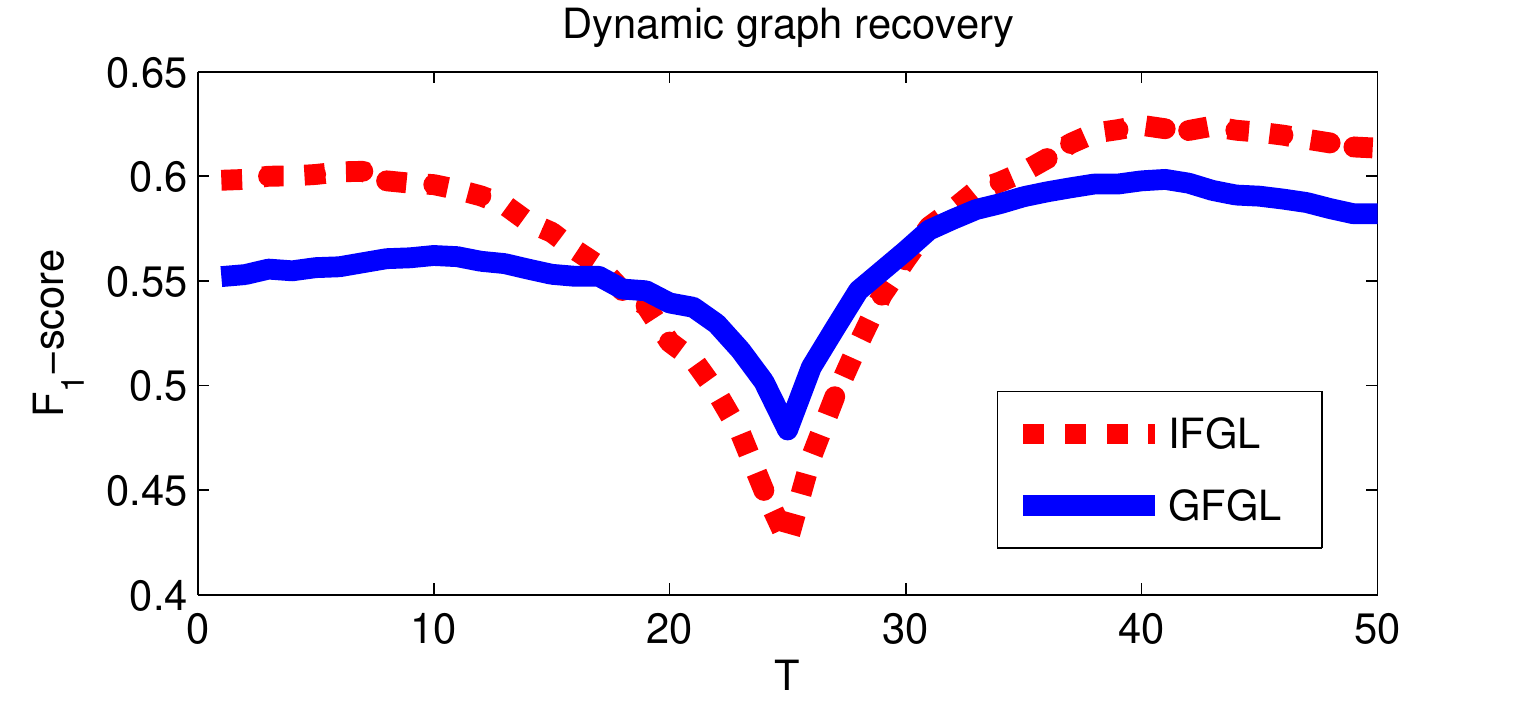}
}
\subfloat[]{
\includegraphics[width=0.5\columnwidth]{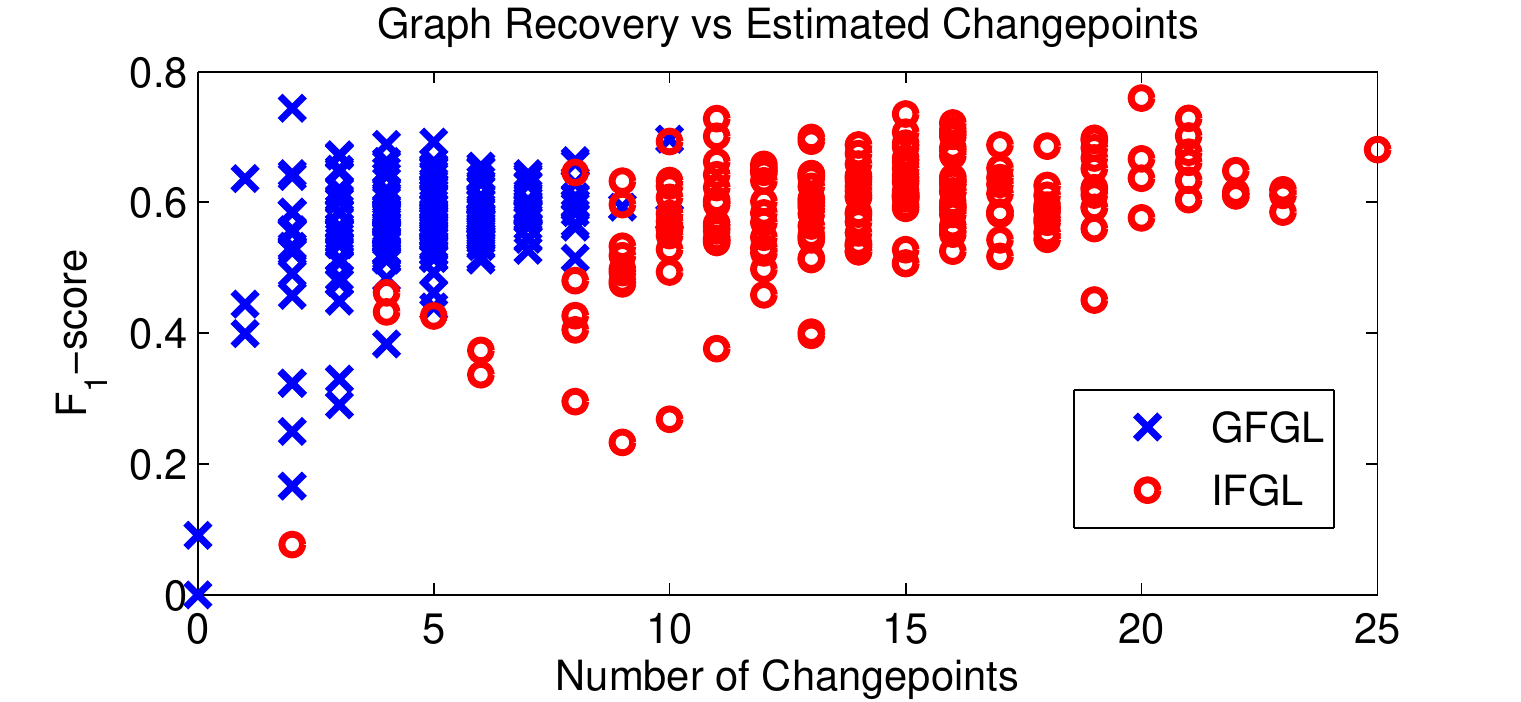}
}
}
\caption{Comparison of generalization performance between GFGL and IFGL, for a 
dataset of size $P=10$, $T=50$. (a) $F_1$-score as a function of 
time $t$, plotted lines
are the averages over $N_{\mathrm{test}}=200$ time-series. Error-bars are
omitted for presentation, with an estimated standard deviation in
the $F_1$-scores of $\sigma_{\mathrm{IFGL}}\approx0.15$ and 
$\sigma_{\mathrm{GFGL}}\approx0.14$.
(b) Demonstration of GFGL and IFGL graph recovery as a function of the number of
estimated changepoints $|\hat{\mathcal{T}}|$.
\label{fig:F-score-(edge-recovery)}}
\end{figure}
For each training
time-series an optimal set of parameters are chosen which maximise the 
$F_{1}$-score,
namely
$\{(\lambda_{1}^{*},\lambda_{2}^{*})_{i}=\arg\max\:
F_{1}(\lambda_{1},\lambda_{2})_{i}\}_{i=1}^{N_{\mathrm{train}}}$. 
The final, learnt optimal parameters
$(\lambda_{1}^{*},\lambda_{2}^{*})$, are computed as the median
value in this training set. A hold-out test set of 
independently simulated time-series is then used to measure the generalization 
performance.
Figure \ref{fig:F-score-(edge-recovery)}a provides a typical comparison
of the graph-recovery ($F_{1}$-score) performance between the IFGL and
GFGL methods throughout the time-series duration.  In this example it
can be seen that IFGL tends to perform best at points far from the
changepoint, whereas GFGL shows a benefit when estimating a graph close
to the changepoint. 
\begin{figure}[h]
\centering{
\subfloat[]{
 \includegraphics[width=0.5\columnwidth]{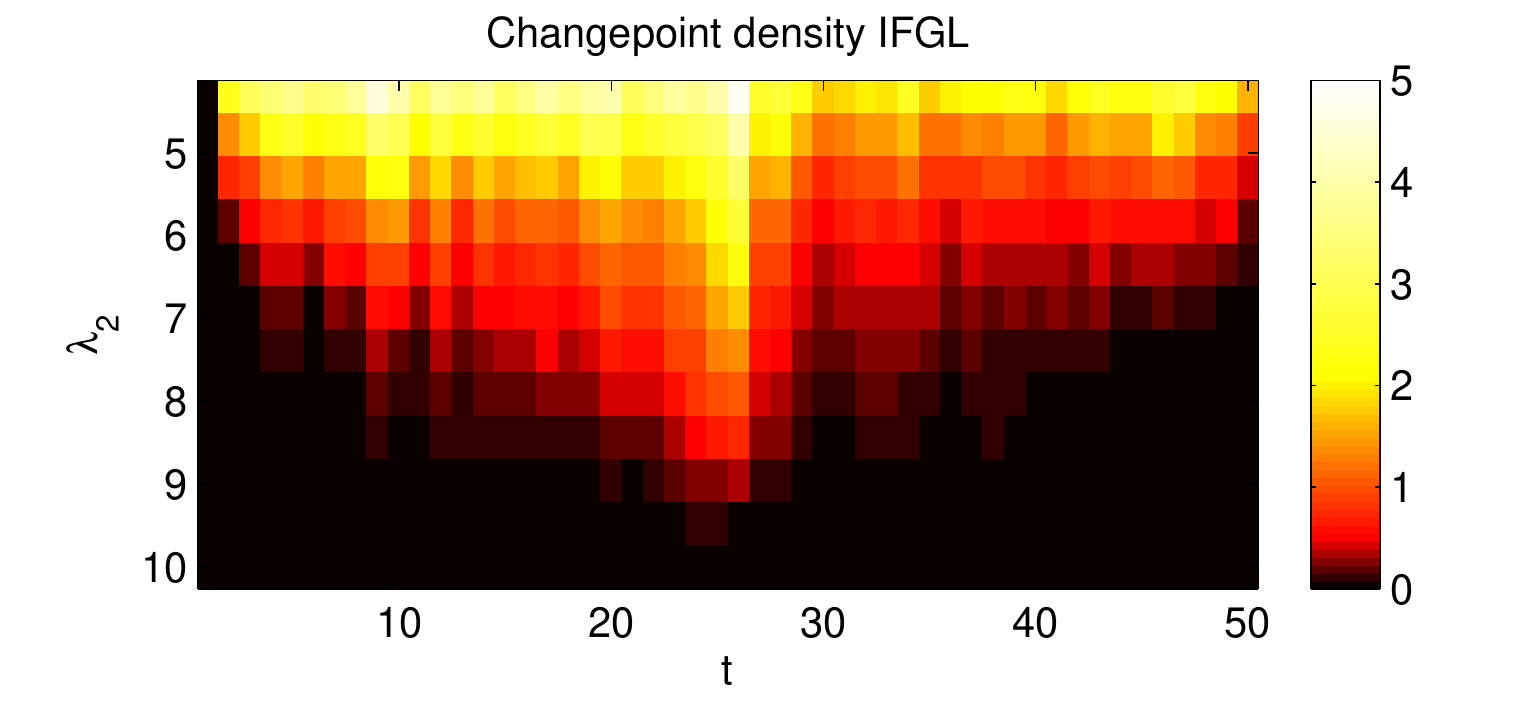}
 }
 \subfloat[]{
 
\includegraphics[width=0.5\columnwidth]{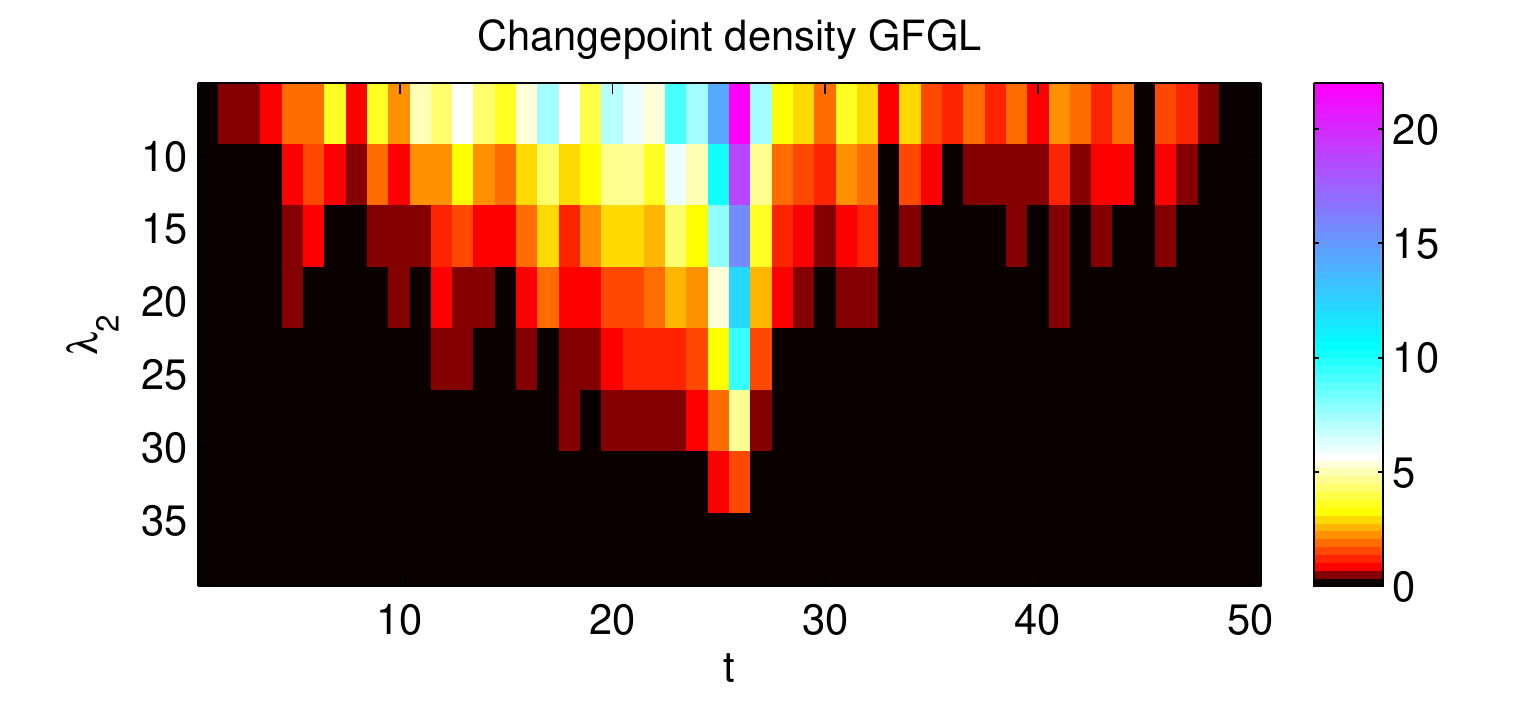}
 }
}
\caption{ Changepoint density plots for IFGL and GFGL in the
synthetic setting ($P=10,\; T=50,\; M=10\; \lambda_1=0.2$), there is a
simulated changepoint at $\mathcal{T}=25$. Color represents the average
number of edges (over $N=100$ simulations) which experience a change at
a given time point.\label{fig:cpd}}
\end{figure}
We note the primary difference between IFGL and GFGL is the number of
edges effected at each changepoint. 
This is demonstrated more clearly in Fig. \ref{fig:cpd}. Here
$\lambda_1$ is fixed and the number of edges which change at each
time-point is plotted over a range of smoothing parameters $\lambda_2$.
Clearly, GFGL results in a greater number of edges being effected at
each changepoint. 
Due to the grouped estimation of GFGL a good level of graph recovery
$F_1$-score performance is achievable with only a few changepoints (see
Figure \ref{fig:F-score-(edge-recovery)}b). In contrast, if one sets
$\lambda_2$ to be large in the IFGL setting, only a few changepoints are
selected; however these represent changes in only very few edges (Fig.
\ref{fig:cpd}a). In this setting IFGL may perform well with regards to
changepoint performance but this comes at the expense of poorer graph
recovery as is evident from the $F_1$-scores. Where such grouped
changepoint structure is present across many edges, GFGL enables one to
recover changepoints without sacrificing as much graphical recovery
performance.

\subsection{Performance scaling}
\label{sub:perf_scaling}
In this section, the recovery performance of the estimators is considered over 
a range of different problem sizes. In order to assess changepoint estimation 
performance and how this varies with
scale, it is insightful to construct an error measure that monitors the average
distance (in time) between estimated and true changepoints.
The changepoints for a given edge $(i,j)$ can be described by
considering differences in the precision matrix, i.e. $
 \hat{\mathcal{T}}_{ij} = 
 \big\{ 
   t \colon 
   \big\vert 
     \hat \Theta_{ij}^t - \hat \Theta_{ij}^{t-1} 
   \big\vert 
   \neq 0, \: t = 2, \ldots, T
 \big\}
 =: 
 \{ \hat \tau_{ij}^k \}_{k=1}^{\hat K_{ij}}
$, 
with $\hat K_{ij} = \big\vert \hat{\mathcal{T}}_{ij} \big\vert$.
These are compared with the ground truth changepoints for the $i,j$th edge
($\tau_{ij}$) from the changepoint set $\mathcal{T}_{ij}$ via the mean absolute
error measure,
namely $
  \text{MAE} := 
  \frac{1}{\hat{K}} 
  \sum_{i,j}
  \sum_{k=1}^{\hat{K}_{ij}}
  \left\vert 
    \hat{\tau}_{ij}^{k} - \tau_{ij}
  \right\vert
$, 
where $\hat K = \sum_{ij} \hat K_{ij}$ \footnote{One should note that
$\hat K = \big\vert \hat{\mathcal{T}} \big\vert$ only when  no
changepoints occur simultaneously across multiple edges; i.e.  
$
 \big\vert \hat{\mathcal{T}} \big\vert 
 =
\hat K
\iff
 \big\vert \bigcup_{ij} \hat{\mathcal{T}}_{ij} \big\vert
 =
 \sum_{ij}
 \big\vert \hat{\mathcal{T}}_{ij} \big\vert 
$.}. 
In these experiments a single changepoint is shared across multiple edges 
 at $\mathcal{T}={T/2}$. To allow fair comparison between
experiments at different time-series lengths ($T$), the same precision matrices
are used either side of the changepoint. For example, under scaling
$T\rightarrow 2T$, the
number of data-points either side of the changepoint is simply doubled. When
considering scaling with respect to dimension precision matrices are
simulated as discussed in Sec. (\ref{sub:sim}); however the number of
active edges scaled as $M=P$. Experiments were run with
data-sets of size $N_{\mathrm{train}}=20$ and $N_{\mathrm{test}}=200$ and optimal
lambdas were selected through $F$-score maximization.
\begin{figure}[h]
\centering{
\subfloat[]{
\includegraphics[width=0.33\columnwidth]{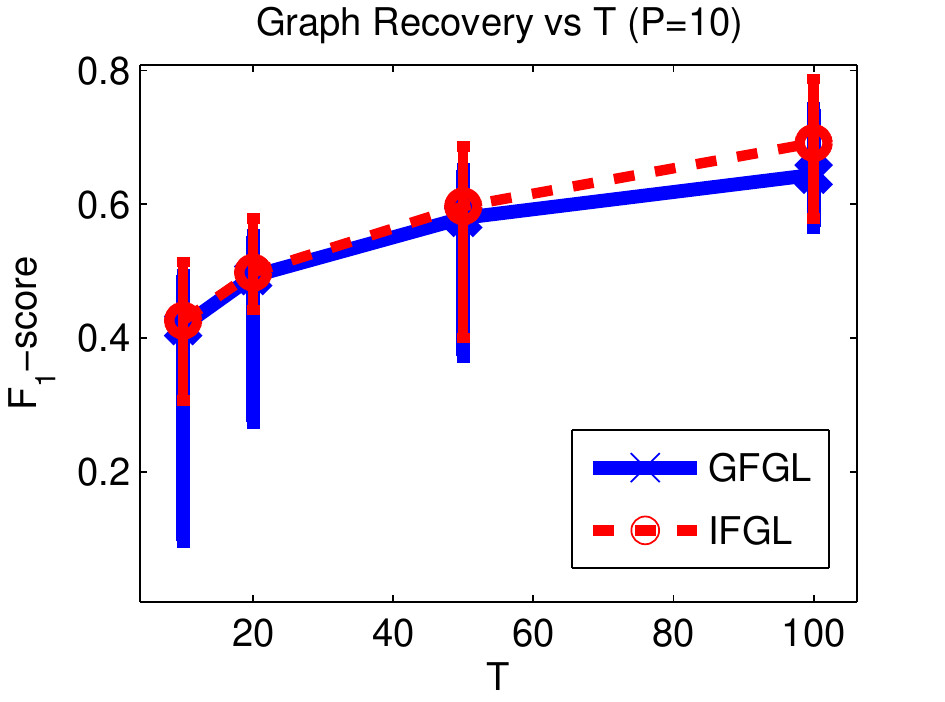}
}
\subfloat[]{
 \includegraphics[width=0.335\columnwidth]
{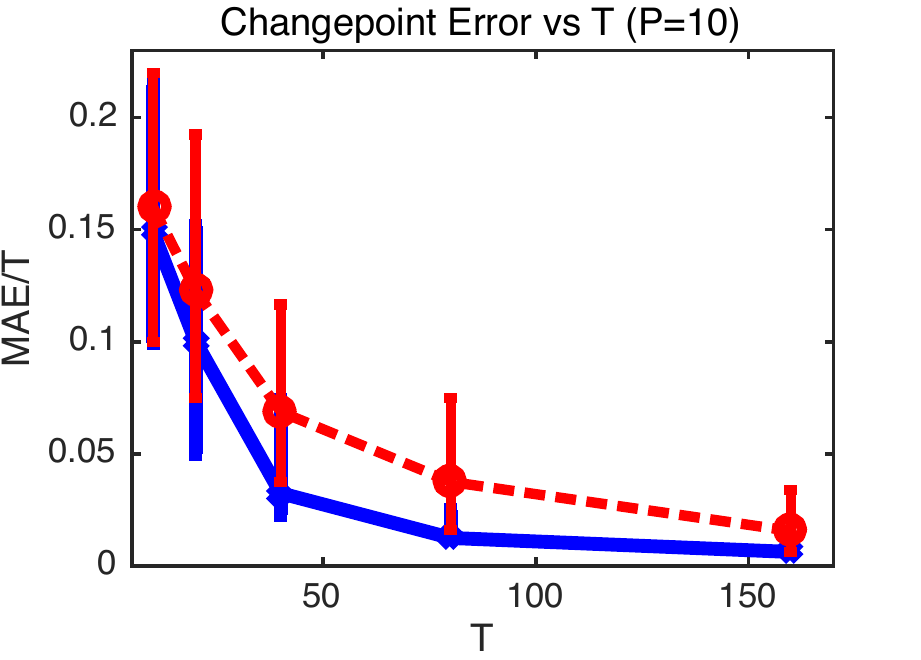}
}
\subfloat[]{
\includegraphics[width=0.33\columnwidth]{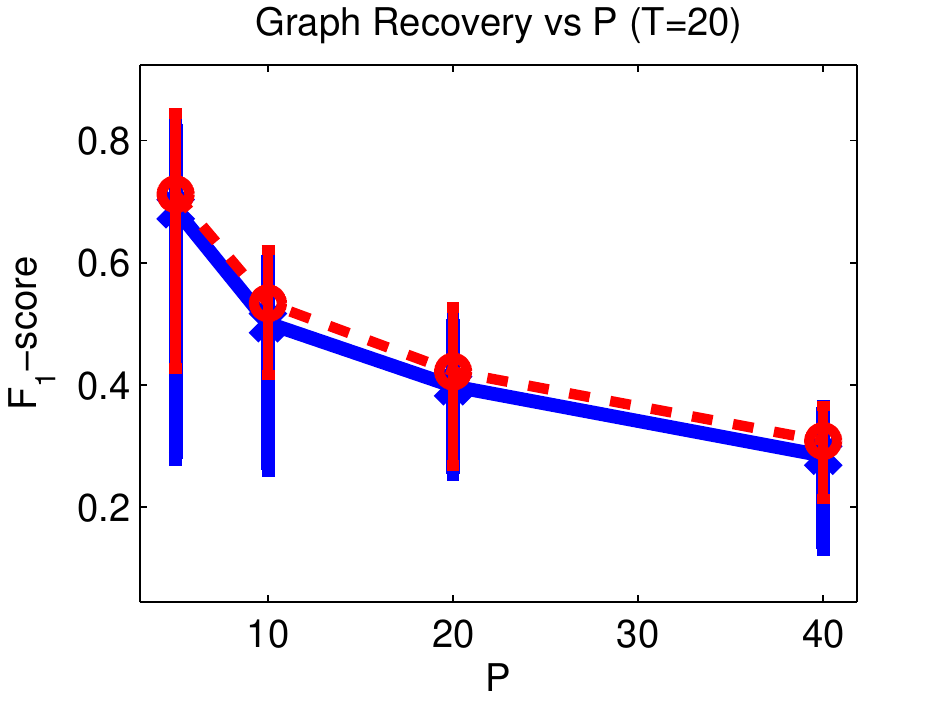}
}
}
\caption{Estimator performance and scaling: (a) $F_{1}$-score vs time-series 
length; (b) relative changepoint error ($MAE/T$) as a function of
increasing time-series length $T$; (c) $F_1$-score vs dimension
$P$. Error bars represent 67\% confidence intervals as estimated from the empirical c.d.f. of $N=200$ test examples.
 \label{fig:Example-of-estimator}
}
\end{figure}
The results presented in Fig. \ref{fig:Example-of-estimator}
demonstrate that recovery performance
improves as more data is made available (increasing $T$) and degrades as the
problem task becomes more
complex (increasing $P$). On average IFGL performs slightly better at
estimating the correct edges. However GFGL performs better in the
changepoint detection task where the relative changepoint error reduces at an improved rate as $T$ 
is increased. Such a result coincides with the performance
demonstrated in Fig. \ref{fig:F-score-(edge-recovery)} where GFGL
outperforms in the vicinity of a changepoint. If grouped changepoints are
present the experiments suggest GFGL performs better in the changepoint 
estimation task without sacrificing graph recovery performance.

  The results here display how recovery performance scales with problem
  dimensionality. However such performance will also depend on the
  structure of the ground-truth graph and precision matrices. As an
  example, in the stationary setting \cite{Ravikumar2011} suggests that,
  for consistent recovery of graphs (with $N$ data-points), one should
  bound the partial correlations, $[-1,-\alpha] \cup [\alpha,1]$, such
  that $\alpha=\Omega(\sqrt{\log P / N})$. To enable better
  interpretation of experimental results we fixed $\alpha=1/2$ in these
  examples. However, it is anticipated that changepoint and graph
  estimation may become more difficult as the true non-zero partial correlations $\Theta_{i,j}$ tend towards zero.

\subsection{Computational Complexity}
\begin{figure}[h]
\centering{
\subfloat[]{\includegraphics[width=0.33\columnwidth]{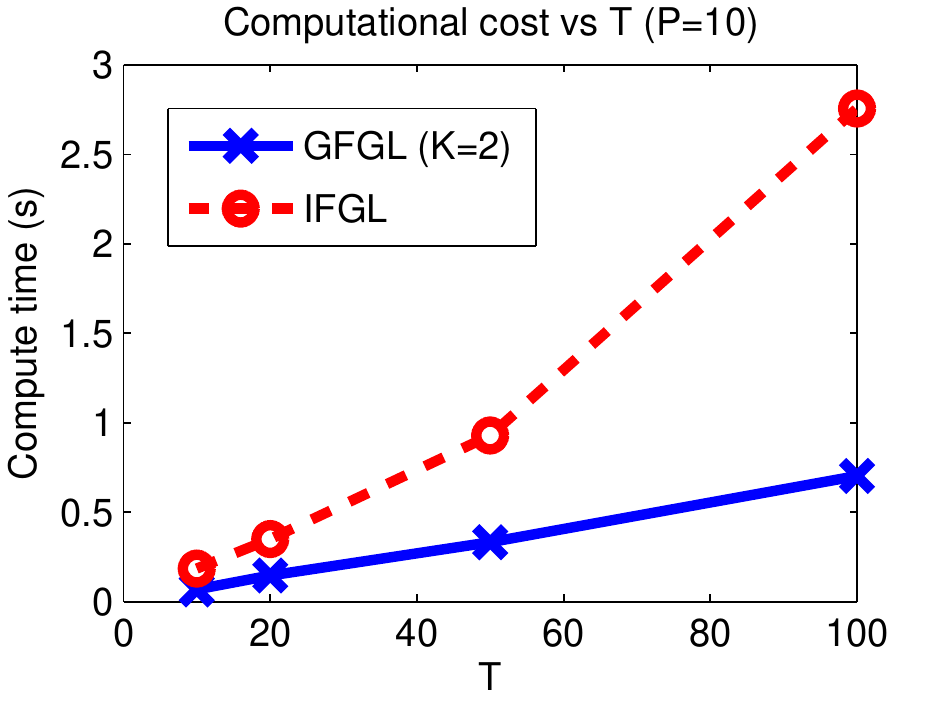}
}
\subfloat[]{
\includegraphics[width=0.33\columnwidth]{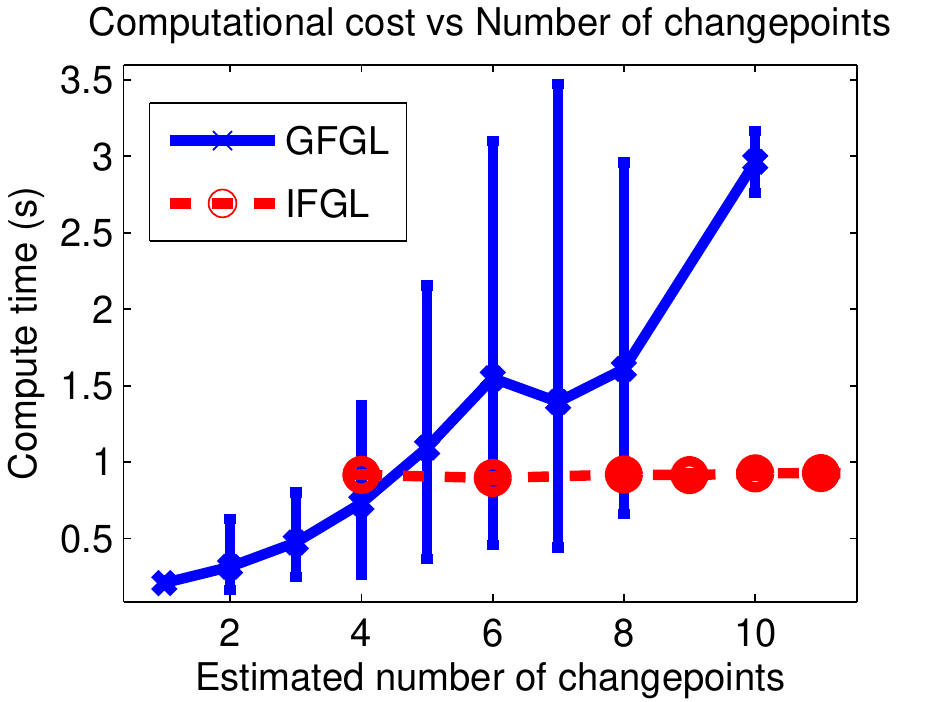}
}
\subfloat[]{
\includegraphics[width=0.33\columnwidth]{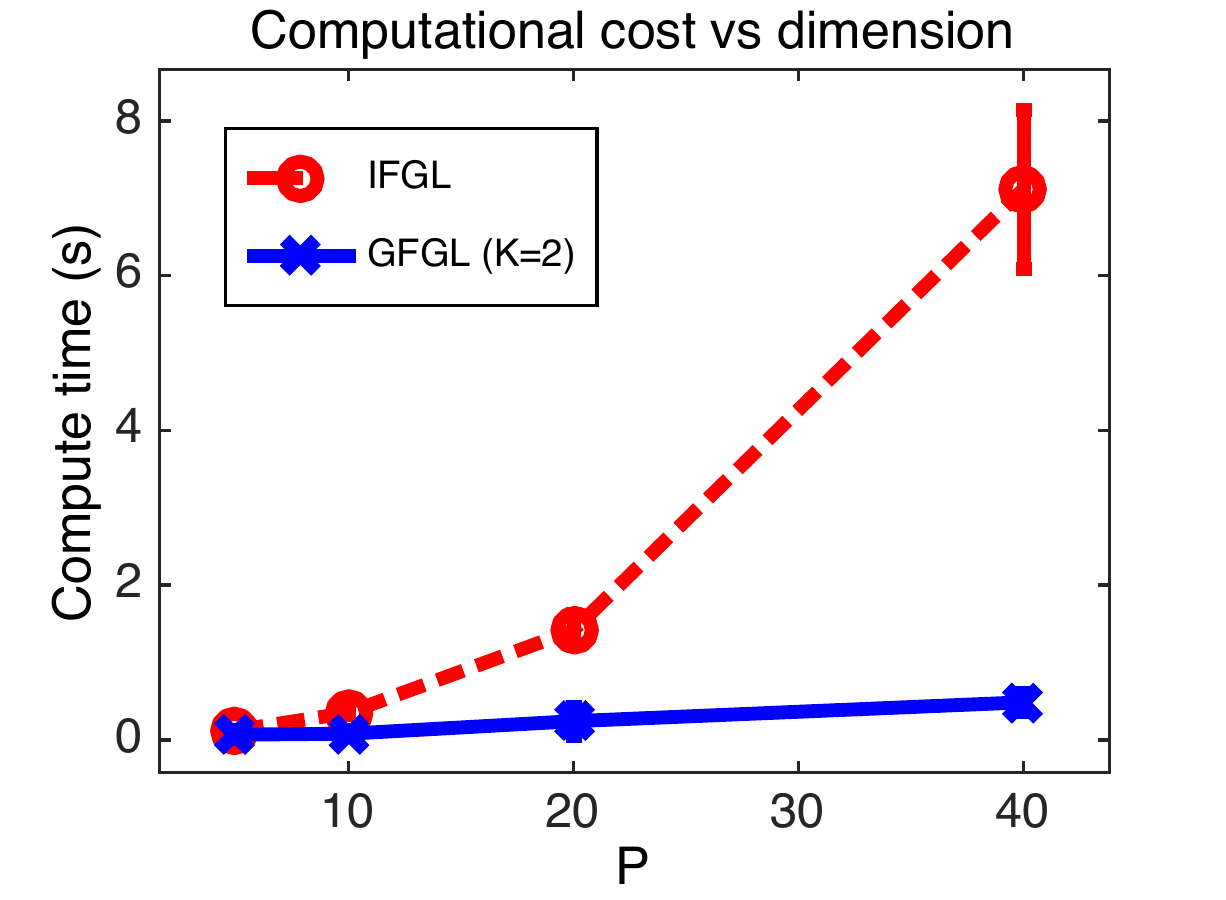}
}
}
\caption{Empirical computational performance of ADMM for GFGL: (a) Compute-time 
vs time-series length $T$ for a fixed number of
changepoints; (b) Compute-time vs number of estimated changepoints 
$\hat{|\mathcal{T}|}$;
(c) Compute-time as a function of dimension $P$.
\label{fig:Timescaling}}
\end{figure}

In order to investigate computational scalability, a series of
experiments were performed on problems of
various size (the experimental setup is the same as in Sec.
\ref{sub:perf_scaling}), the results are summarized in Figure
\ref{fig:Timescaling}. 
In contrast with the quadratic time complexity for
dynamic programming methods \citep{Angelosante2011}, it can be observed that
the ADMM routine, as a whole, maintains roughly linear complexity with
increasing $T$. When considering increases in the estimated number of
changepoints $\hat{K}$, complexity appears to follow the quadratic rate of
GFLseg (used in Algorithm \ref{al:ADMM}), which scales as
$\approx\mathcal{O}(TP^2\hat{K}^{2})$, see 
\cite{Bleakley}.

\section{Example: Time Evolution of Genetic Dependency Networks}
\label{sec:Applications}
In this section we give an example of how methods such as GFGL can be used in an applied context. In recent years it has become increasingly common to construct
experiments which sample gene-expression activity as a time-series. 
As an 
example of such data, we consider the genetic activity of a fruit-fly
(\emph{D.  melanogaster}) from its embryonic birth to final adult state.
The dataset we analyze is a subset of the data collected by
\cite{Arbeitman2002}, which measures gene expression patterns for 4096 genes,
approximately one third of all \emph{D. melanogaster} genes, over $T=67$ time-points.

To aid interpretation of the results and for computational feasibility,
we consider a smaller subset of genes ($P=150$), which are understood to
be linked to certain biological processes, in this case, immune system
response. The link between this subset of genes and biological function
is motivated by considering conserved \emph{co-domains} of a gene. Where
such co-domains are shared between genes, one can often infer a similar
biological function of the genes, this similarity can be extended to
other organisms if the genes are homologous \citep{Forslund2011}. In
this case, our selection of genes is based on the Flybase
\citep{flybase2016} Gene-ontology database. Understanding the dependency
between genes involved in a certain process is interesting to biologists
who want to examine and understand why or how regulation of gene
activity evolves over time--- for example after an intervention or treatment.
Previous work on this data-set by \cite{Lebre2010} considered estimating changepoints in a causal VAR-type model. In contrast to this work, we are concerned with estimating the contemporaneous relationships between genes. Specifically, we model the innovations $\boldsymbol{\epsilon}_{t} $, where $
 Y_{t} = Y_{t-1} + \boldsymbol{\epsilon}_{t} \;
 \mathrm{where} \; 
 \boldsymbol{\epsilon}_{t} \sim \mathcal{N} 
 (\boldsymbol{0}, \boldsymbol{\Theta}^{t}) \;.
$

\begin{figure}[h]
\centering{
\includegraphics[height=6.2cm]{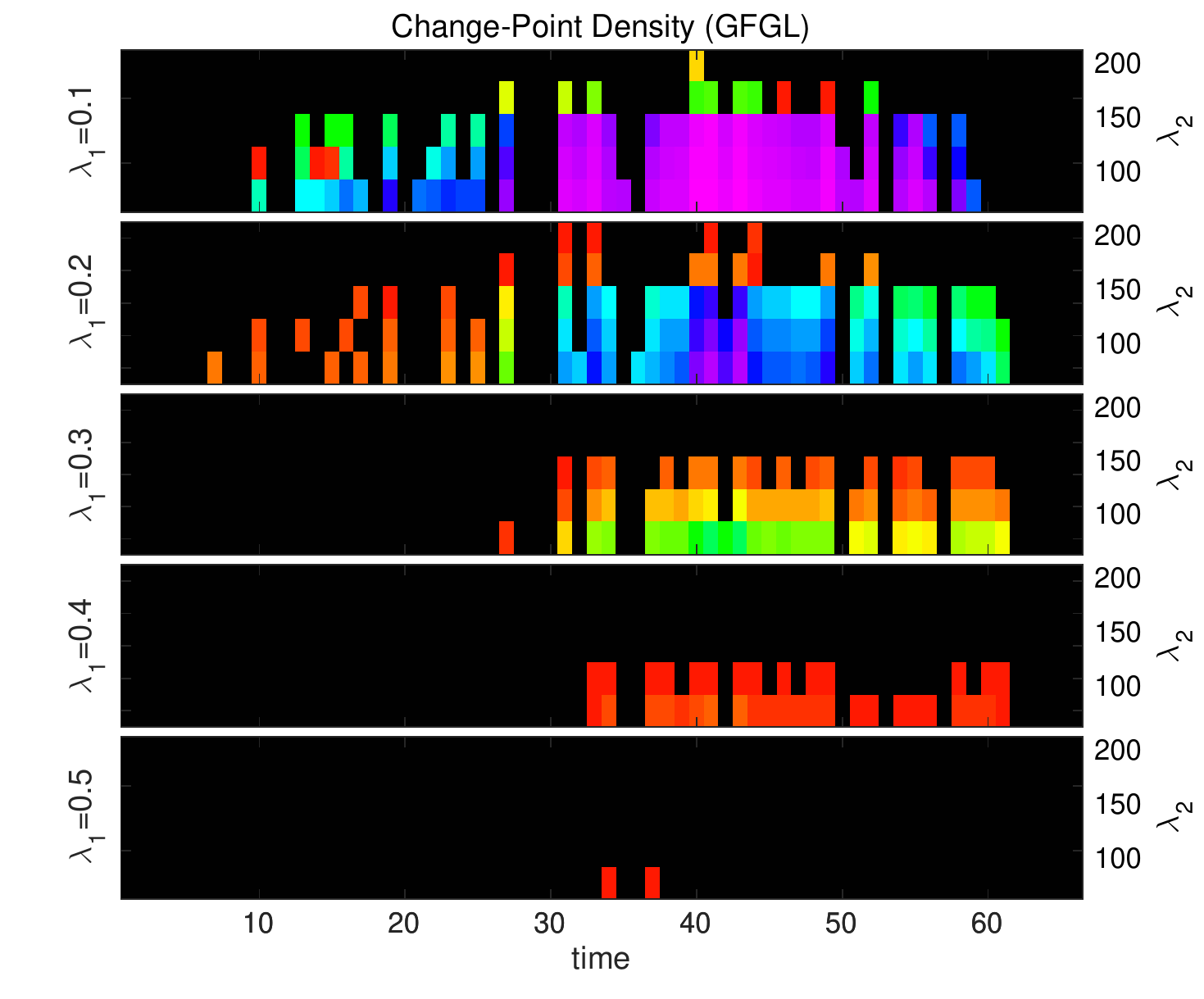}
\includegraphics[height=6.2cm]{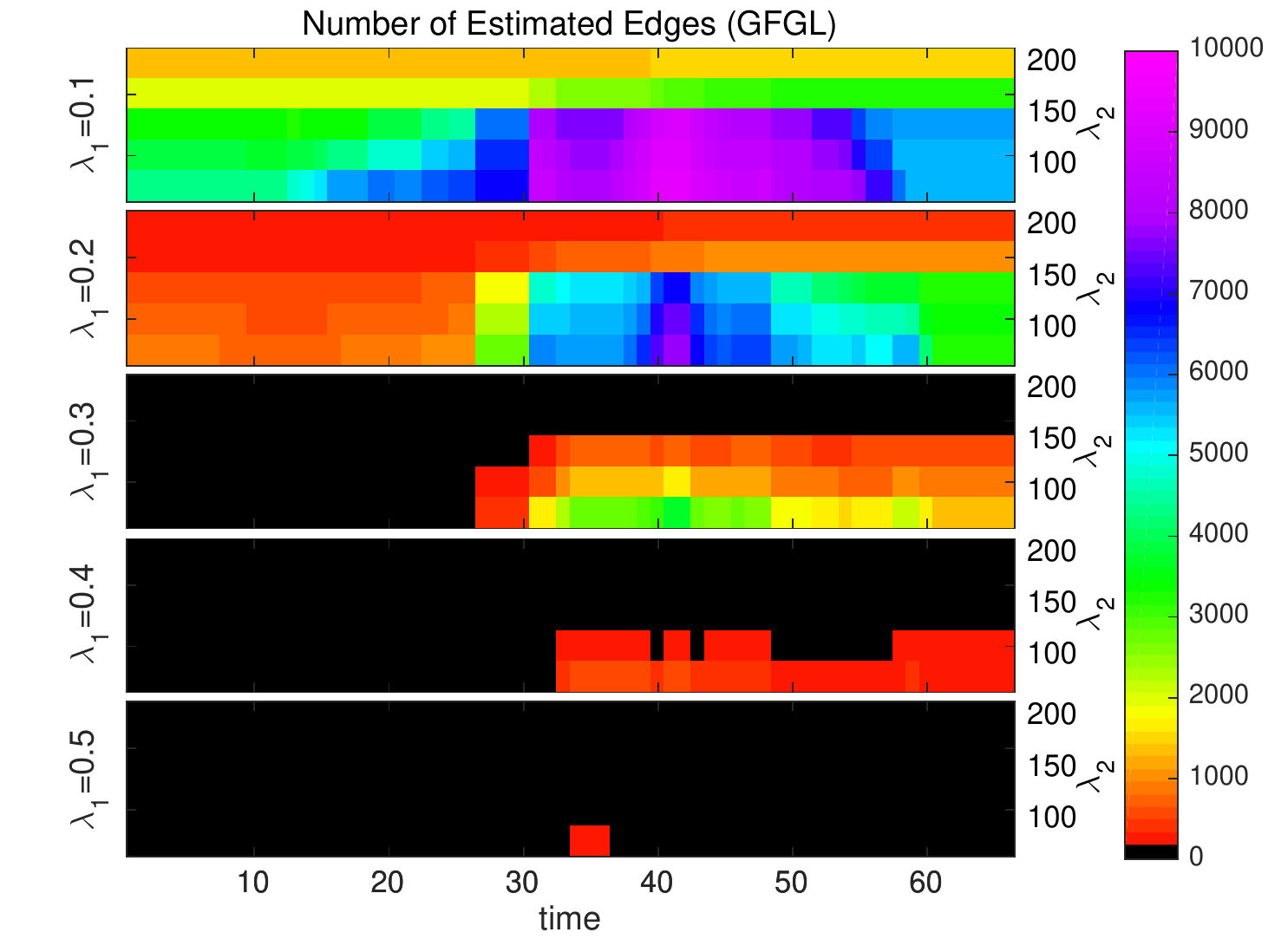}\\
\includegraphics[height=6.2cm]{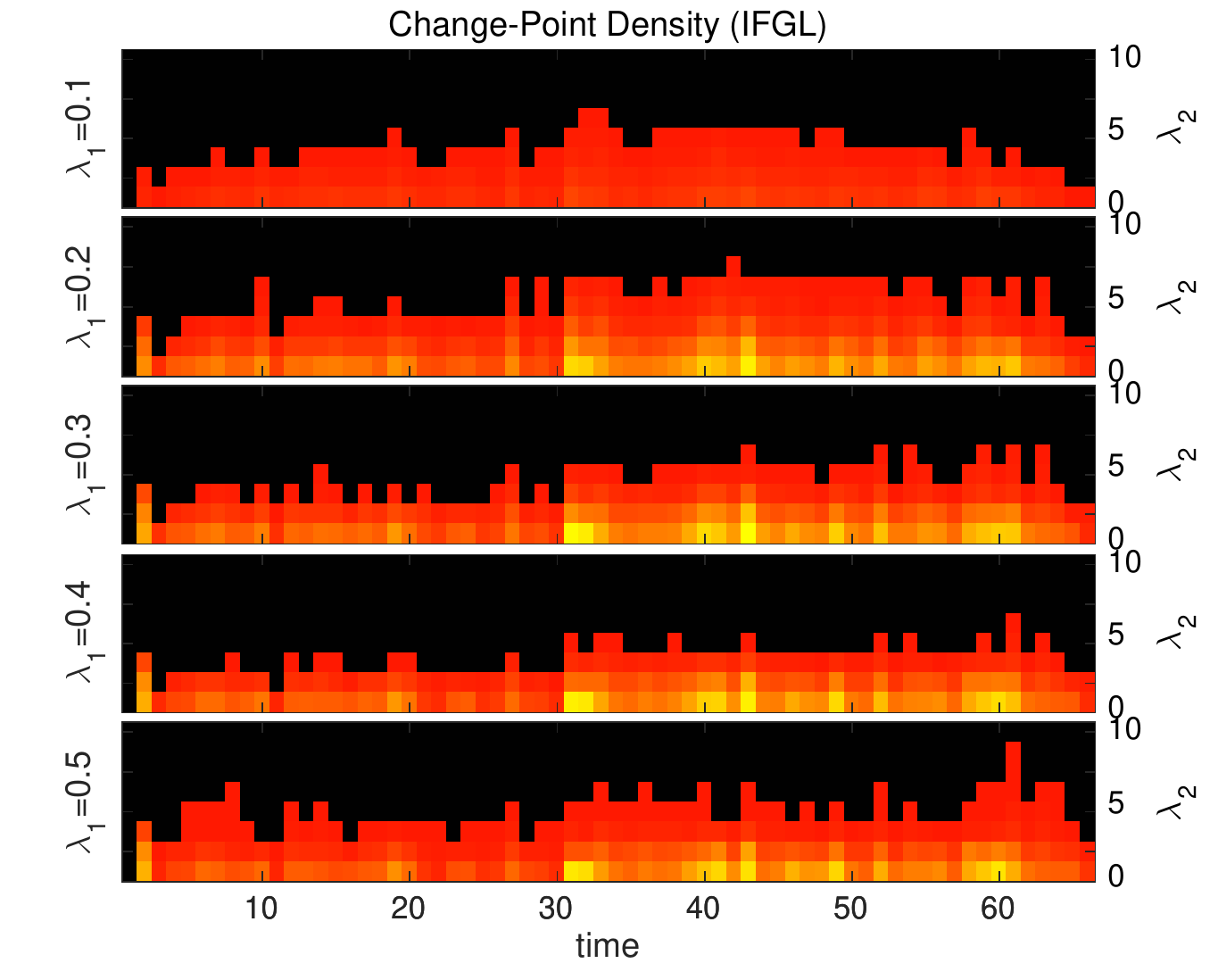}
\includegraphics[height=6.2cm]{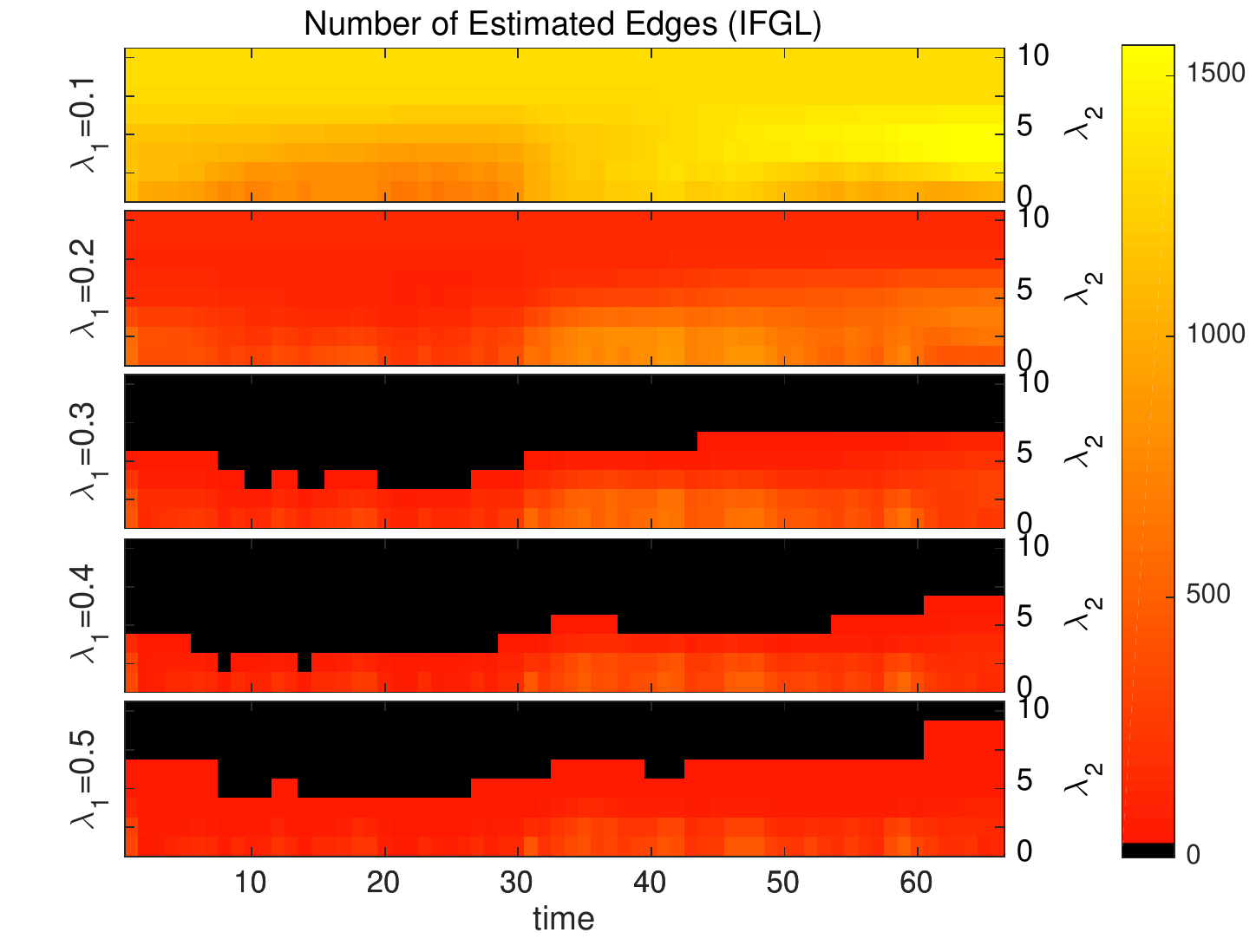}
}
\caption{Changepoint density (the color indicates the number of edges which change) and the number of edges recovered as function of both time and tuning parameters.}
\label{fig:CPD_GFGL}
\end{figure}

Unlike in the synthetic experiments, the time-course data analyzed here was not replicated, i.e. we only have one data-point at each time point in the fly's development. It is worth noting that more recent experiments involving time-course microarray data may produce replicated experiments. These are thought to be particularly valuable, as it allows one to gauge the uncertainty due to variation in genetic populations and environmental factors. With such replicated experiments, there may also be a meaningful way to perform cross-validation to estimate the hyper-parameters.
However, in the absence of replicates, we adopt an exploratory approach and consider the inferred structure over a wide range of regularization parameters. We scan over the range $\lambda_1=0.1$ to $0.5$ for both methods, with $\lambda_2=80$ to $200$ for GFGL, and $\lambda_2=1$ to $10$ in the IFGL case.

\begin{figure}[h!]
\centering{
\includegraphics[width=\columnwidth]{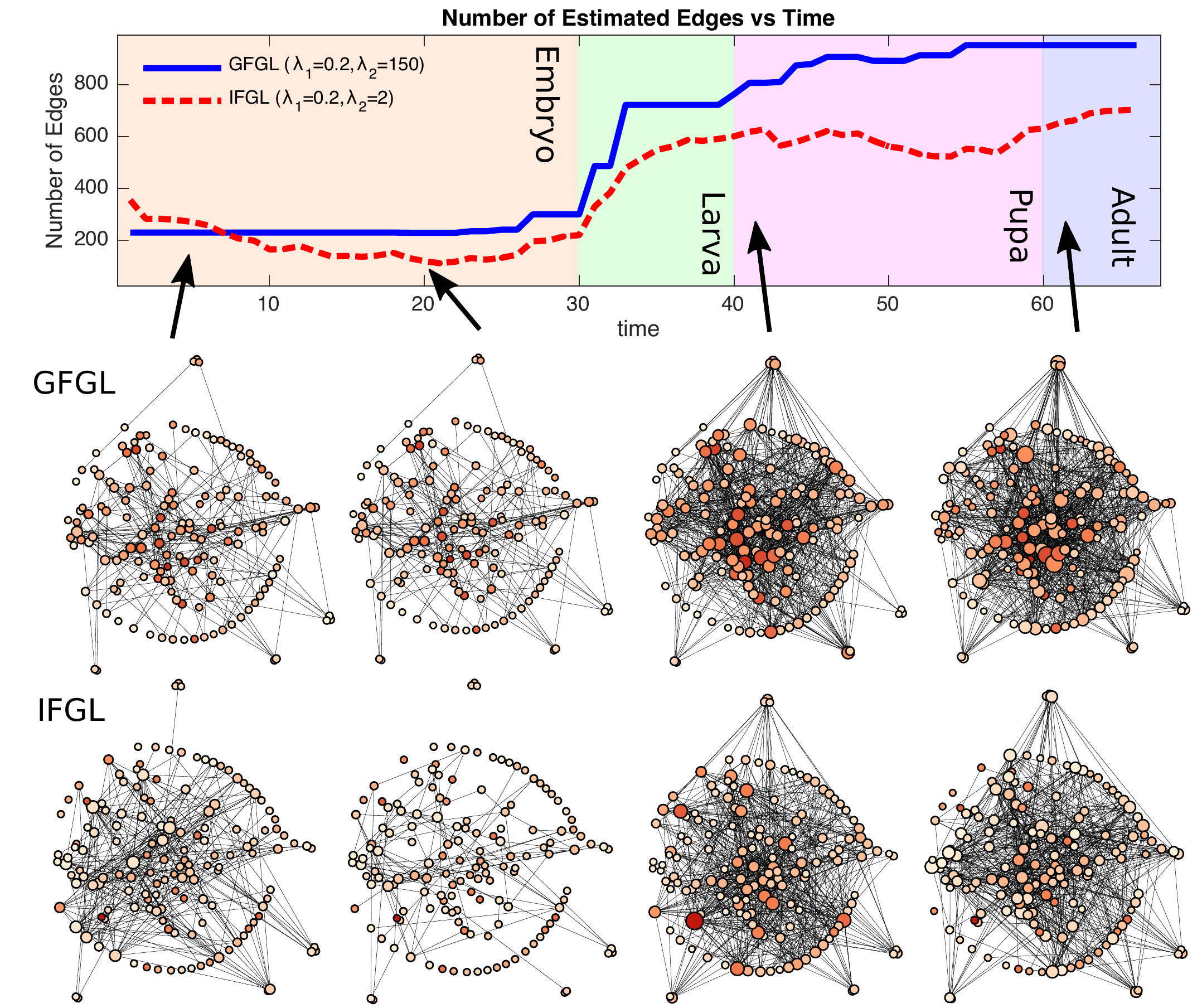}
}
\caption{Top: Estimated Edges as a function of time with overlay of physiological life-cycle stages. Bottom: Estimated graph structure for GFGL and IFGL at 4 different time-points ($t=5,t=20,t=40,t=62$), size of node indicates degree, the positions of nodes (representing individual genes) are comparable across graphs.}
\label{fig:Example_graphs}
\end{figure}

 Figure \ref{fig:CPD_GFGL} demonstrates how both the sparsity, number
 and position of changepoints in the solution behave as a function of
 $\lambda_1,\lambda_2$. One can clearly see that both smoothing (the
 number of changepoints) and sparsity (the number of edges) are linked
 to $(\lambda_1,\lambda_2)$ jointly.  For a given selection of
 $\lambda_1,\lambda_2$ we obtain an estimate of the dynamic graph, some
 snapshots of such graphs are illustrated in Fig.
 \ref{fig:Example_graphs}. 
 In this example, the graphs are drawn such
 that gene-positions (vertices) are comparable both across time and
 between methods. This application to genetic data clearly illustrates
 the qualitative differences between the estimators in terms of
 extracted structure. In both methods we observe that more edges are
 detected in the later-half of the life-cycle, with a
 large change in structure inferred during the Larval stage of development. Unlike
 IFGL, which experiences \emph{changepoints} at all time-points, GFGL
 clearly has more pronounced jumps; i.e. more edges change at each
 changepoint (see Fig. \ref{fig:CPD_GFGL}). Additionally, if one
 considers the varying size of node (proportional to degree) it
 appears that the degree of the GFGL estimates are more stable. Such a
 feature suggests that the particular GFGL estimate (in Fig.
 \ref{fig:Example_graphs}) has fewer degrees of freedom than the IFGL
 estimate. Such a property may be appealing in the high-dimensional
 setting, where GFGL appears to permit similar graphical structure, but
 with enhanced temporal stability in the graph.

\section{Discussion}
\label{sec:Discussion}

Two classes of estimators have been investigated for piecewise constant
GGM. In particular, we have proposed the GFGL estimator for grouped
estimation of changepoints in a dynamic GGM. Empirical results suggest
that GFGL has similar model recovery abilities to the IFGL class of
estimators.
However, when simultaneous grouped changepoints are expected to occur,
the group-fused estimator does not appear to sacrifice as much
graph-recovery performance in order to accurately estimate changepoints. Further to this,
when estimating grouped changepoints, the group-fused estimator appears to converge to the true changepoints at a faster rate.
We find that the grouped approach offers a more meaningful and
interpretable segmentation of the graphical dynamics. This is especially
apparent when such grouped changes represent systemic phase or regime
changes in activity.
When one has a priori knowledge of grouping, it is anticipated that GFGL
will offer a useful and scalable investigative tool to support data
exploration and subsequent inference.

Our empirical results on the relative changepoint error and F-score vs
$T$ suggest convergence of the estimator.  However, we leave further theoretical examination
of such consitency properties as future work. A possible way forward here is to exploit
the theory developed for the individual subproblems, namely changepoint
detection with the lasso \citep{Harchaoui2010} and sparse group lasso \citep{Zhang,Simon2013}. Such work may build on results in the dynamic graph learning setting by \cite{Kolar2011,Kolar2012} and \cite{Roy2015}.

\section{Supplemental Materials}

An Appendix containing technical details and further information can be found in the on-line supplemental material for this paper. In addition to the Appendix, one can also find a compressed folder with example code to implement the GFGL and IFGL estimators as discussed in the paper.

\bibliographystyle{jasa}

\bibliography{GFGL_final}

\newpage

\end{document}